\documentclass[final]{statsoc} 
\pdfoutput=1

\usepackage[utf8]{inputenc}
\usepackage[a4paper]{geometry}
\usepackage[title]{appendix}
\usepackage{amsmath}
\usepackage{amssymb}
\usepackage{bm}
\usepackage{hyperref}
\usepackage{graphicx}
\usepackage{subfigure}
\usepackage{algorithm}
\usepackage{algpseudocode}
\usepackage{etoolbox}
\usepackage[dvipsnames]{xcolor}
\usepackage[normalem]{ulem}
\usepackage{multirow}
\usepackage{etoolbox}
\usepackage[dvipsnames]{xcolor}
\usepackage{ulem}
\usepackage{xr}

\counterwithout{equation}{section}

\DeclareMathOperator{\diag}{diag}

\newcommand{\bs}{\boldsymbol}

\makeatletter
\newcommand*{\addFileDependency}[1]{
  \typeout{(#1)}
  \@addtofilelist{#1}
  \IfFileExists{#1}{}{\typeout{No file #1.}}
}
\makeatother

\newcommand*{\myexternaldocument}[1]{%
    \externaldocument{#1}%
    \addFileDependency{#1.tex}%
    \addFileDependency{#1.aux}%
}

\myexternaldocument{supp_mat}

\title[Survival modelling in earthquake early warning]{Survival modelling of smartphone trigger data for earthquake parameter estimation in early warning. With applications to 2023 Turkish-Syrian and 2019 Ridgecrest events}

\author{Luca Aiello}
\address{Department of Economics, Management and Statistics, Università degli Studi di Milano-Bicocca, Italy.}
\author{Raffaele Argiento}
\address{Department of Economics, Università degli Studi di Bergamo, Italy}
\author{Francesco Finazzi}
\address{Department of Economics, Università degli Studi di Bergamo, Italy}
\author[Aiello, Argiento, Finazzi and Paci]{Lucia Paci}
\address{Department of Statistical Sciences, Università Cattolica del Sacro Cuore, Italy}
\email{francesco.finazzi@unibg.it}

\begin{document}

\begin{abstract}
    Crowdsourced smartphone-based earthquake early warning systems recently emerged as reliable 
    alternatives to the more expensive solutions based on scientific-grade instruments.
    For instance, during the 2023 Turkish-Syrian deadly
    event, the system implemented by the Earthquake Network citizen science initiative provided a forewarning up to 25 seconds. 
    We develop a statistical methodology based on a survival mixture cure model which provides full Bayesian inference on epicentre, depth and origin time, and we design an efficient tempering MCMC algorithm to address multi-modality of the posterior distribution. 
    The methodology is applied to data collected by the Earthquake Network, including  the 2023 Turkish-Syrian and 2019 Ridgecrest events.
\end{abstract}

\keywords{cure models; mixture modelling; Bayesian analysis; Markov chain Monte Carlo; citizen science.}

\section{Introduction}\label{sec:introduction}

Earthquake early warning systems (EEWSs) \citep{gasparini2007earthquake} are deployed in seismic-prone areas for mitigating the earthquake risk. Dense networks of instruments are used to detect earthquakes in real-time, to stop critical processes and to send alerts to people who will experience strong ground shaking. Construction and operating costs of classic EEWSs are in the order of the millions of euros \citep{given2014technical}. This limited their widespread, especially in low-income seismic countries that could largely benefit from EEWSs.

In the last two decades, researchers have proposed low-cost alternatives \citep{cochran2009quake, clayton2011community}. Only with the smartphones revolution, however, new solutions \citep{finazzi2016earthquake, kong2016myshake} became truly accessible to the global population. The Earthquake Network (EQN) citizen science initiative \citep{finazzi2020earthquake} was the first implementing a smartphone-based EEWS able to send real-time alerts on people's smartphones. Since 2013, more than 8 million people took part into the initiative and the system contributed to send nearly 6,000 alerts. During the 2023 Turkish-Syrian earthquake, EQN provided a forewarning up to 25 seconds to people exposed to very-high levels of ground shaking, proving its efficiency in earthquake risk mitigation.

With respect to scientific-grade instruments used in classic EEWSs, smartphones are less reliable since they are subject to anthropic noise and the earthquake monitoring is based on a low-cost accelerometer. Thus, specific statistical methodologies have been developed for the real-time detection of earthquakes \citep{finazzi2017statistical,massoda2023} and for assessing the detection performance of the smartphone-based monitoring network \citep{finazzi2022probabilistic}. 

In this work, following the same line, we extend and improve the above works developing a statistical methodology for estimating relevant earthquake parameters which are epicentre, depth and origin time. The methodology deals with specific peculiarities of smartphones when used as seismic sensors, namely the possibility to detect accelerations unrelated with earthquakes, the possibility to miss an earthquake and the uncertainty on which seismic wave is detected by the smartphone.

We embed the estimation problem into the framework of survival data analysis by exploiting the analogy between right censored clinical data, which are naturally analysed with survival models, and smartphone-based earthquake data. In particular, smartphones are seen as patients of a clinical study. A smartphone which detects accelerations (triggering smartphone) possibly induced by the earthquake is seen as a patient dying, while a smartphone that does not detect any acceleration is viewed as a patient surviving. The earthquake plays the same role of  a disease, so that smartphones that will never trigger are interpreted as \emph{cured} patients. The earthquake detection by the EQN system is then the censoring event. Indeed, the triggering times of (uncured) smartphones that will eventually trigger after the earthquake detection are not observed.

In population-based clinical studies, \emph{cure} occurs when the hazard rate in the diseased group of individuals returns to the same level as that expected in the general population \citep{peng2014cure}. In other words, a certain proportion of subjects in the population is not expected to experience the events of interest and so the relative survival curve appears to plateau after a given time. Mixture survival modelling has been widely employed to incorporate this plateau and estimate the cure fraction in clinical studies. For instance, \citet{deangelis1999} applied a  mixture  survival model to analyse individual data on colon cancer and estimate both the proportion of cured patients and the failure time distribution for fatal patients. \citet{lambert2007} compared the estimates of relative survival and the cure fraction between mixture and non-mixture cure fraction models. \citet{lambert2010estimating} extended the mixture cure model by incorporating a finite mixture of parametric distributions to provide flexibility in the shape of the relative survival or excess mortality functions. A Bayesian mixture cure model with spatial random effects has been proposed by \citet{yu2012} to model county-level cancer survival data using Markov chain Monte Carlo (MCMC)  methods, while \citet{lazaro2020} implemented  a Bayesian  mixture cure model using the integrated nested Laplace approximation.

 Summing up, the contribution of this work is to propose a Bayesian cure model for analysing smartphone-based early warning data in order to estimate relevant earthquake parameters, uncertainty included. The model leverages information from spatial locations and times of triggering smartphones as well as from spatial locations of non-triggering smartphones. As a key feature, the mortality function (i.e., the smartphone triggering function) is assumed as a mixture of two parametric densities describing the earthquake waves that may be detected by a smartphone in a given time interval.  
On the computational side, an efficient sampling strategy has been designed to provide full posterior inference of the cure fraction and the earthquake parameters. Specifically, we develop an adaptive parallel tempering MCMC algorithm to address the computational challenges associated with the multi-modality of the posterior distribution. 

Through a simulation study and a set of relevant case studies we illustrate the soundness and the good performance of our modelling approach. Specifically, the approach is applied to the EQN detections of the magnitude 7.8 2023 Turkish-Syarian earthquake, of the magnitude 7.1 2019 Ridgecrest (US) earthquake and of a relatively small magnitude earthquake occurred in Mexico with epicentre offshore, thus more challenging in terms of epicentre estimation.

The rest of the paper is organised as follows. Section~\ref{sec:EQN} describes the functioning of EQN system and the dataset generated when an earthquake is detected, including the three datasets analysed in this work.
Section~\ref{sec:model} develops the Bayesian mixture cure  model and describes how the physics of the earthquake is exploited 
in the model. Computational details are then presented in Section~\ref{sec:comp_det}.  Section~\ref{sec:simulation} offers the results of a simulation study while Section~\ref{sec:case_study} illustrates our methodology with three relevant case studies.
Finally, Section~\ref{sec:discussion} provides a brief summary and discusses some future works.

\section{EQN functioning and data generation}\label{sec:EQN}

This section provides an overview on the functioning of the EQN system and on the dataset generated every time an earthquake is detected. EQN deploys a global smartphone network for seismic monitoring. To join the EQN initiative, smartphone users must install the EQN app which is available for Android, iOS and Huawei operating systems. Seismic monitoring starts when the smartphone is charging and not used. This implies that the network is highly dynamic, with smartphones entering and leaving the network at any time. A central server collects the monitoring data from the smartphones and it knows the state of the network, namely which smartphones are monitoring and where they are located. On each smartphone, seismic monitoring is done through the smartphone accelerometer which continuously measures and provides the smartphone acceleration in space. A monitoring smartphone possibly detects a seismic wave but also any other acceleration induced by a local force (e.g., someone moving the smartphone or the object above which the smartphone is located). 

Earthquakes are characterised by primary (P) and secondary (S) waves. The P is faster than the S wave and it usually produces a mild ground shaking. Most of the earthquake damages are caused by the S wave. When an earthquake hits, the monitoring smartphone may detect the P or the S wave depending on the earthquake magnitude, on the distance between the smartphone and the epicentre, on where the smartphone is located (e.g., on which floor of the building and above which object) and on the specific accelerometer sensitivity (which may differ from a smartphone to another). The monitoring smartphone may also not detect the earthquake, either because the epicentre is too far or because the smartphone is not able to send the information to the server.

When the server detects an earthquake using the algorithm detailed in \cite{finazzi2017statistical}, two lists of data are stored: the list of smartphones which detected an acceleration (triggering smartphones) and the list of smartphones which were known to be monitoring at the detection time but did not trigger (active smartphones). Triggering smartphones are described by their spatial coordinates and by their triggering times, while active smartphones are only described by their spatial coordinates. The list of triggering smartphones covers the period that goes from two minutes before the EQN detection time to detection time.

Both lists cover an area of $300$ km in radius centred on the EQN detection location, which is provided by the detection algorithm as a rough estimate of the earthquake epicentre. This estimate, however, assumes that all smartphones triggered because of the earthquake, it does not take into account the information of the non-triggering smartphones and it does not consider the dynamic of the P and S waves. It follows that the EQN epicentre estimate is usually largely biased.

\subsection{EQN datasets}\label{sec:EQN_datasets}

To better understand their complexity and before entering into modelling details, we describe here the three EQN datasets analysed in Section \ref{sec:case_study}.
Table~\ref{tab:cs_details} provides, for the three earthquakes, the number of triggering smartphones, the number of active smartphones that did not trigger, the hypocentre coordinates, the origin time and the earthquake magnitude.

\begin{table}
    \caption{\label{tab:cs_details} EQN datasets summaries and parameters of the three earthquakes analysed in this work. Earthquake parameters were retrieved on February 16, 2023 from the \href{https://www.emsc-csem.org}{European-Mediterranean Seismological Centre}.}
    \centering
        \begin{tabular}{l|rrr}
        & Turkey-Syria & Ridgecrest & Mexico \\ \hline
        EQN triggers & 16 & 80 & 150 \\ 
        EQN active smartphones & 158 & 241 & 1115 \\ 
        Latitude [deg] & 37.17 & 35.76 & 15.64 \\ 
        Longitude [deg] & 37.08 & -117.62 & -94.97 \\ 
        Depth [km] & 20.0 & 8.0 & 27.9 \\ 
        Origin time [UTC] & 2023-02-06  & 2019-07-06  & 2019-07-17  \\ 
        & 01:17:36 & 03:19:52 & 06:25:48 \\ 
        Magnitude [M\textsubscript{W}] & 7.8 & 7.1 & 4.8
        \end{tabular}
\end{table}

The first dataset is related to the recent magnitude 7.8 Turkish-Syrian earthquake of February 6, 2023. The event was the deadliest and strongest earthquake occurred in Turkey since 1939, and the second most powerful after the 1668 North Anatolia earthquake. Additionally, it was the deadliest earthquake in modern-day Syria. EQN detected the event with a delay of 10 seconds with respect to the earthquake origin time, allowing people living far away from the epicentre to receive a forewarning. Panel (a) of Figure~\ref{fig:summary_turkey} depicts the EQN dataset generated when the earthquake was detected. From a visual analysis of the triggers and of the active smartphones, it is easy to delimit the area not yet impacted by the earthquake.

The second dataset is about the magnitude 7.1 Ridgecrest (US) earthquake of July 6, 2019. This was the most powerful earthquake occurred in the California in 20 years and was largely felt across the state and in Nevada. Due to the small number of smartphones in the epicentral area at the time of the event, the EQN detection occurred in Los Angeles at around 190 km (see \citealt{bossu2022shaking} for details), limiting the EQN early warning capabilities but still providing real-time information to citizens.  
Panel (a) of Figure~\ref{fig:summary_california} shows the EQN dataset. Smartphones are mainly located around the biggest cities of the study region, i.e., Los Angeles and San Diego. Moreover, triggering times look noisy and they do not necessarily increase with the distance from the earthquake epicentre. 

The third dataset concerns a magnitude 4.8 earthquake occurred on July 17, 2019 offshore the Mexican coast. The event had no impact on people but it is useful to assess our methodology when the epicentre is ``outside" and far from the smartphone network. EQN detected the earthquake at around 90 km from the epicentre, thanks to the smartphones located in Salina Cruz and Juchitán de Zaragoza (the two cities along the coast closest to the epicentre). Again, panel (a) of Figure~\ref{fig:summary_mexico} shows the EQN dataset. Despite the earthquake is detected on the coast, the dataset includes many triggers from areas not yet reached by the seismic waves. EQN is very popular in Mexico and each city is a source of spurious triggers which tend to bias the epicentre estimate. 

The features of the three datasets calls for a statistical model that exploits all the available information (including the information of the non-triggering smartphones), and that takes into account the dynamic of seismic waves.

\section{Model developments}
\label{sec:model}

\subsection{Mixture cure model for relative survival}

As anticipated in the \nameref{sec:introduction}, our inferential problem can be formalised within the setting of survival data analysis with right censoring and cure. Let $T$ be the smartphone triggering time and let $T^{*}$ be the censoring time, equal to the time at which the earthquake is detected by EQN. Hence, we denote $Y =\min{(T, T^{*})}$, $y$ its realisation and $\delta = I(T<T^{*})$ the censoring indicator. In other words, $\delta = 1$ when the smartphone triggered.

As usual in survival analysis, the survival function gives the probability that a subject (a smartphone) will survive (will not trigger) past time $t$, i.e., $S(t) = P(Y > t) = 1 - F(t)$. In particular, in relative survival models \citep{deangelis1999,lambert2010estimating}, the overall (all-cause) survival function $S(t)$ is given by the product of the expected survival function $S_{0}(t)$ and the relative survival function $R(t)$, that is
\begin{equation}\label{eq:surv}
    S(t) = S_{0}(t)R(t).
\end{equation}
 The relative survival function $R(t)$ can be interpreted as the net survival probability, i.e., the survival probability, when the disease (earthquake) is the only cause of death (triggering) and all other causes could be eliminated. In our setting, $S_{0}(t)$ represents the expected probability that a smartphone will not trigger until time $t$ for any other causes other than the earthquake. 

Given Equation \eqref{eq:surv}, the overall hazard function $h(t)$ is obtained as the sum of two components: the expected hazard function $h_{0}(t)$ and the excess hazard function $\lambda(t)$, that is
\begin{equation}\label{eq:haz}
    h(t) = h_{0}(t) + \lambda(t).
\end{equation}
The assumption behind Equations \eqref{eq:surv} and \eqref{eq:haz} is that the triggering associated with the quake is independent of triggering associated with other causes, i.e., there are independent competing risks \citep{gamel2001non}.
Both $S_0(t)$ and $h_0(t)$ in Equation \eqref{eq:haz} are assumed known and equal, respectively, to  the survival and hazard function of an exponential distribution with  parameter $1/86400$ seconds$^{-1}$, assuming that  the probability of non-triggering due to the earthquake decays over time at the  rate of one expected trigger per day.

Customary, as time goes to infinity, the survival curve goes to 0, i.e., $S(\infty)=0$. However, in our context, this assumption is not realistic since, in practice, a smartphone could never trigger because of the earthquake. Here, cure models come in. Cure models refer to a class of models for censored survival data from subjects, when some of them will not develop the event of interest, however long they are followed. The subjects who are not going to develop the event of interest are often referred to as \emph{cured} subjects or long-term survivors. See \citet{peng2014cure} for a review on these models.

Since some smartphones will never trigger, it is natural to consider a mixture model where the population is a mixture of two groups: (i) a proportion $\pi$ of cured cases with respect to the specific cause (smartphones that will never trigger because of the earthquake), and thus having a similar mortality rate to that expected in the general population (i.e., smartphones that will eventually trigger for other causes), and (ii) a proportion $(1-\pi)$ of uncured cases, i.e., smartphones that will eventually trigger for the earthquake whose survival function will tend to zero. Let W be the cure indicator with $W = 1$ if the smartphone is cured, where $P(W = 1) = \pi$. Define $S_{Q}(t) = P(T > t \mid W = 0)$ the survival function of the uncured population, i.e., the survival function of smartphones that will eventually trigger because of the earthquake. The mixture cure model is then defined by the following survival function
\begin{equation}
    S(t) = S_{0}(t)\{\pi + (1-\pi)S_{Q}(t)\},
\end{equation}
where the survival function regarding the $\pi$ component is equal to one since a cured smartphone has probability one not to trigger up to  $t$, i.e., it will never trigger because of the earthquake. Note that the cure definition encoded in the parameter $\pi$ is from a population perspective and it does not provide any information on individual smartphones. The corresponding hazard rate can be expressed as the sum of the background triggering rate and the triggering rate associated with the earthquake, that is
\begin{equation}
    h(t) = h_{0}(t) + \frac{(1-\pi)f_{Q}(t)}{\pi + (1-\pi)S_{Q}(t)},
    \label{eq:ht}
\end{equation}
where $f_{Q}(t)$ is the probability density function associated with $S_{Q}(t)$. Hence, in relative cure survival models, the relative survival function has an asymptote at the cure fraction $\pi$ or, equivalently, the excess hazard function $\lambda(t)$ has an asymptote at zero.

The density $f_{Q}(t)$ in Equation \eqref{eq:ht} is modelled to capture earthquakes dynamic. In particular, we express the density $f_{Q}(t)$ as a mixture of two densities corresponding to the P and S waves, that is $f_{Q}(t) = \alpha f_{P}(t) + (1-\alpha) f_{S}(t)$, where the definition/the choice of $f_P(t)$ and $f_S(t)$ is detailed in the next Section~\ref{sec:fitting}. Hence, the survival function becomes
\begin{equation}\label{eq:S}
    S(t) = S_{0}(t)\{\pi + (1-\pi)(\alpha S_{P}(t) + (1-\alpha) S_{S}(t))\},
\end{equation}
with hazard function
\begin{equation}
    h(t) = h_{0}(t) + \frac{(1-\pi)(\alpha f_{P}(t) + (1-\alpha) f_{S}(t))}{\pi + (1-\pi)(\alpha S_{P}(t) + (1-\alpha) S_{S}(t))}.
    \label{eq:hazard}
\end{equation}
Similar to \citet{lambert2010estimating},  this model can be thought of as a three-component mixture: (i) a proportion $\pi$ of cured individuals, i.e., smartphones will never trigger because of the earthquake, (ii) a proportion $(1 - \pi)\alpha$ of uncured individuals with survival distribution $S_{P}(t)$, i.e., the proportion of smartphones detecting the P wave and (iii) the remaining proportion $(1-\pi)(1-\alpha)$ of uncured individuals with survival distribution $S_{S}(t)$, i.e., the proportion of smartphones detecting the S wave.

Denote $\bs{\theta} = (\mathbf{z}_{0},d_{0},t_{0})$ the earthquake parameter we aim at estimate, i.e., the coordinates of the epicentre, $\mathbf{z}_{0} = (lon_{0},lat_{0})$, the earthquake depth, $d_{0}$, and the event origin time, $t_{0}$. Given $n$ active smartphones in the study area over the observational window, let  $y_{i}$, $i=1,\dots,n$, be the minimum between the time associated to the signal sent by smartphone $i$ and the EQN detection time $t^\ast$ and $\mathbf{z}_{i}=(lon_{i},lat_{i})$ the GPS coordinates of such device. If $y_{i}$ is uncensored ($\delta_{i} = 1$), the $i$-th smartphone contributes $f(y_{i}| \bs{\theta}, \mathbf{z}_i)$, to the likelihood while if $y_{i}$ is censored ($\delta_{i}  = 0$), the $i$-th smartphone contributes $S(y_{i}\mid \bs{\theta}, \mathbf{z}_i)$ to the likelihood, namely
\begin{equation}\label{eq:likelihood}
    \mathcal{L}(\bs{\theta},\pi,\alpha;\mathbf{y}, \mathbf{Z}, \bs{\delta}) = \prod_{i=1}^n f(y_{i}\mid \bs{\theta}, \mathbf{z}_i)^{\delta_{i}} S(y_{i}\mid \bs{\theta}, \mathbf{z}_i)^{1-\delta_{i}} = \prod_{i=1}^n h(y_{i}\mid \bs{\theta}, \mathbf{z}_i)^{\delta_{i}} S(y_{i}\mid \bs{\theta}, \mathbf{z}_i),
\end{equation}
where $\mathbf{y}=(y_1, \dots, y_n)$, $\mathbf{Z}=(\mathbf{z}_1, \dots, \mathbf{z}_n)$ and $\boldsymbol{\delta}=(\delta_1\dots, \delta_n)$.
Moreover, $h(y_{i}\mid \bs{\theta}, \mathbf{z}_i)$ and $S(y_{i}\mid \bs{\theta}, \mathbf{z}_i)$ are the hazard and survival functions defined in Equations \eqref{eq:S} and \eqref{eq:hazard} where we  explicitly denote the dependence on $\bs{\theta}$ and $z_i$.

\subsection{P and S waves density functions}
\label{sec:fitting}
As we already mentioned, earthquakes are characterised by primary (P) and secondary (S) waves. The P wave travels at around $7.8$ km/s and usually causes little damage. The S travels at around $4.5$ km/s and is the most energetic wave. Smartphones may detected the P or the S wave depending on the shaking intensity caused by each wave at the smartphone location. 

Given the earthquake origin time $t_0$, the hypocentre ($\mathbf{z}_{0},d_{0}$) and the smartphone coordinates, the expected times at which P and S waves reach the device are easily computed. Due to the functioning of the detection algorithm implemented on the smartphone and to Internet latency, the triggering may occur within $3.5$ seconds from the wave arrival. Hence, a possible choice for $f_{P}(y_{i}\mid\bs{\theta},\mathbf{z}_{i})$ and $f_{S}(y_{i}\mid\bs{\theta},\mathbf{z}_{i})$ are uniform distributions over the $3.5$ seconds temporal support where the triggering may occur. The left bounds of the supports for the P and S waves are defined as follows
\begin{equation*}
    \begin{aligned}
        a_{P} &= \frac{dist_{H}(\mathbf{z}_{i},\mathbf{z_{0}},d_{0})}{v_{P}} + t_{0}\\  a_{S} &= \frac{dist_{H}(\mathbf{z}_{i},\mathbf{z_{0}},d_{0})}{v_{S}} + t_{0},
    \end{aligned}
\end{equation*}
where $v_{P}$ and $v_{S}$ are the P and S wave velocities, respectively, and $dist_{H}(\mathbf{z}_{i},\mathbf{z_{0}},d_{0})$ is the Euclidean distance between smartphone $i$ and the hypocentre $(\mathbf{z}_{0},d_{0})$. 

\begin{figure}[htbp!]
    \centering
    \begin{tabular}{ll}
        (a) & (b) \\
        \includegraphics[width = 0.49\textwidth]{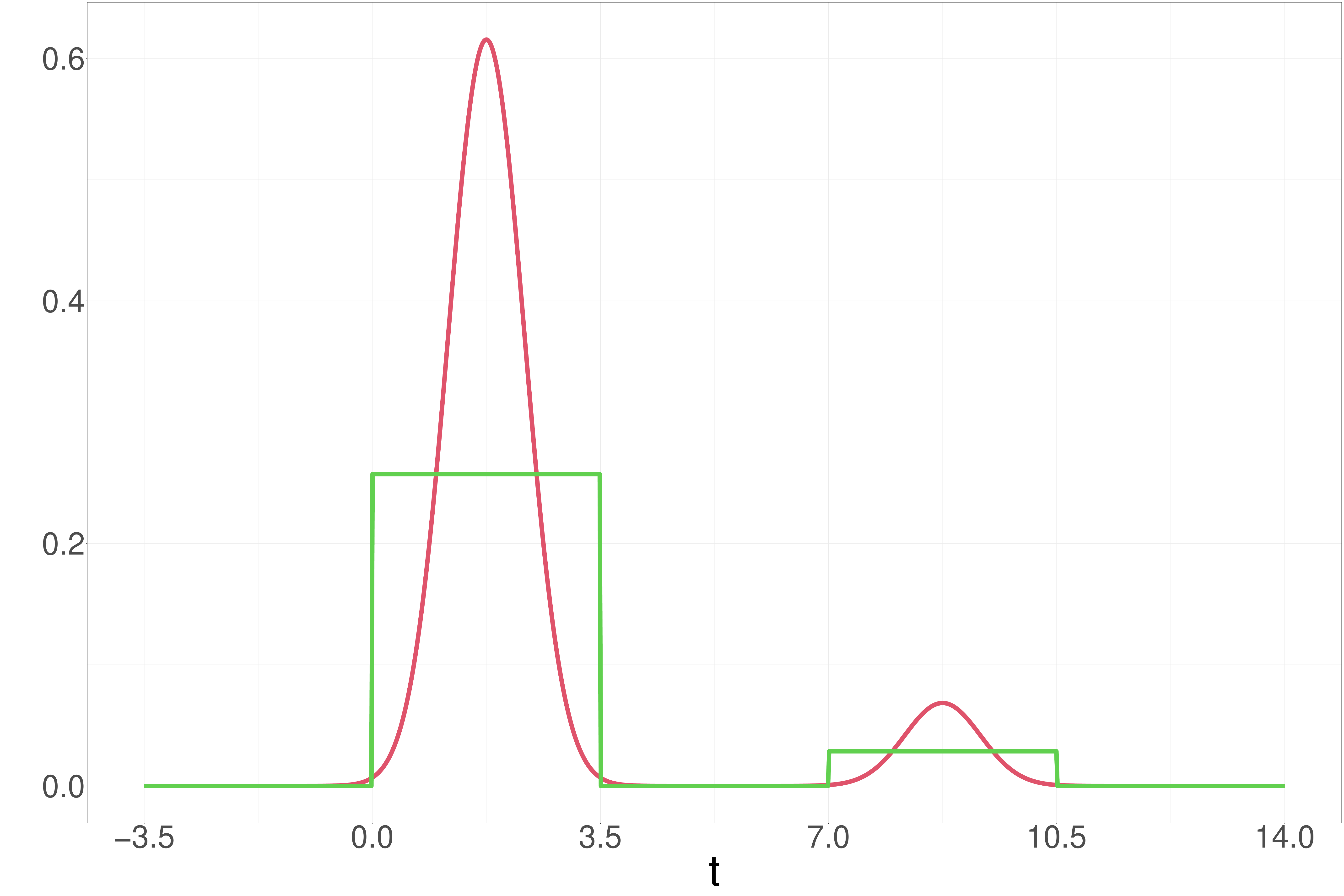}
    &
        \includegraphics[width = 0.49\textwidth]{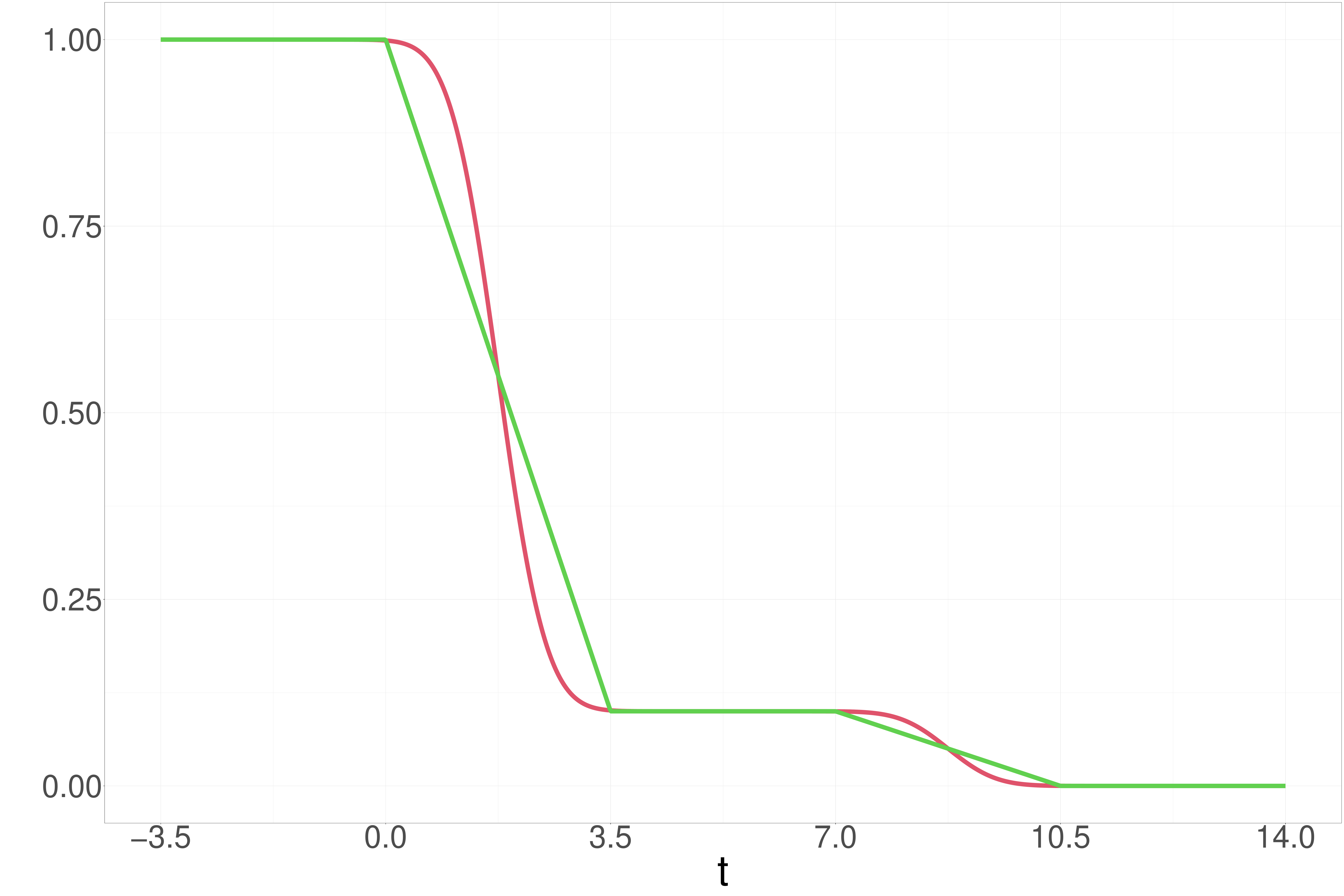}
    \end{tabular}
    \caption{Uniform (green) vs Gaussian (red) mixture with $0.9$ and $0.1$ weights of the P and S waves, respectively. Probability density function (a) and survival function (b).}
    \label{fig:mixture_components}
\end{figure}

Unfortunately, uniform densities pose numerical challenges since the likelihood function in Equation \eqref{eq:likelihood} becomes discontinuous. Thus, any algorithm to explore the posterior parameter space will struggle to find sharp and reliable solutions. For this reason, we assume $f_{P}(y_{i}\mid\bs{\theta},\mathbf{z}_{i})$ and $f_{S}(y_{i}\mid\bs{\theta},\mathbf{z}_{i})$ as the densities of a normal distribution with mean $\mu_{P}(\bs{\theta},\mathbf{z}_{i})$ and $\mu_{S}(\bs{\theta},\mathbf{z}_{i})$, respectively, and variance $\tau^2$, where
\begin{equation}\label{eq:gauss_mixt_par}
    \begin{aligned}
        \mu_{P}(\boldsymbol{\theta},\mathbf{z}_{i}) &= \frac{dist_{H}(\mathbf{z}_{i},\mathbf{z_{0}},d_{0})}{v_{P}} + t_{0} + \mu\\ 
        \mu_{S}(\boldsymbol{\theta},\mathbf{z}_{i}) &= \frac{dist_{H}(\mathbf{z}_{i},\mathbf{z_{0}},d_{0})}{v_{S}} + t_{0} + \mu.\\ 
    \end{aligned}
\end{equation}
The values $\mu$ and $\tau^2$ are chosen in order to have $99\%$ of the probability  of the normal mixture in the $3.5$ seconds interval defined by the uniform mixture. As an example, Figure~\ref{fig:mixture_components} shows the uniform (green) and the Gaussian (red) mixture with weights equal to $0.9$ and $0.1$ for the P and S waves, respectively. Panel (a) displays the density functions while panel (b) shows the corresponding survival functions. Hence, the choice of the values of $\mu$ and $\tau^2$ allows the survival function of the normal mixture to be the closest continuous function to the survival of the uniform mixture.

We refer the survival model \eqref{eq:S}-\eqref{eq:hazard} together with the P and S waves mixture density to as the Survival EarthQuake Model (SEQM).

\subsection{Prior distributions}\label{sec:prior}

The Bayesian hierarchy of the SEQM is completed by specifying the prior distributions for all of the model parameters. Let $\mathbf{z}^{\ast} = (lon^{\ast},lat^{\ast})$ be the EQN detection location,  which is a rough estimate of the earthquake epicentre computed as the triggers centre of mass. For the epicentre coordinates $\mathbf{z}_{0}$, we place a bivariate normal distribution centred on $\mathbf{z}^{\ast}$ with diagonal covariance matrix, i.e., $\mathbf{z}_{0}\sim \mbox{N}\left(\mathbf{z}^{\ast}, \sigma^2 I_{2}\right)$, where $\sigma=1$. For the depth parameter $d_{0}$, we specify a uniform distribution with support between 0 and 100 km, i.e., $d_{0}\sim \text{Uniform}(0, 100)$. For the origin time $t_{0}$, an exponential distribution is assumed, i.e., $t_0 \sim \mbox{Exp}(\lambda)$. The rate $\lambda$ is set to $1/20 \text{ seconds}^{-1}$ since we expect origin time and EQN detection time to be close. As far as the wave mixing weight $\alpha$ is concerned, we place a  $\mbox{Beta}(a, b)$ distribution where $a=1/2$ and $b=1/2$ in order to assume, a priori, that only one wave is detected by the smartphone network. Finally, a uniform between zero and one is assumed on the cure fraction $\pi$, i.e., $\pi \sim \mbox{Uniform}(0,1)$.

\section{Computational details}\label{sec:comp_det}

To approximate the joint posterior distribution under the SEQM with prior described in Section~\ref{sec:prior}, we design a MCMC sampling strategy based on an adaptive parallel tempering Metropolis-within-Gibbs sampler. 

We write the model parameters vector as $\bs{\phi} =(\bs{\phi}_{1},\bs{\phi}_{2},\bs{\phi}_{3}) = (\bs{\theta},\alpha,\pi)$, where, we recall that $\bs{\theta} =(\mathbf{z}_{0},d_{0},t_{0})$. The three blocks of parameters are a priori independent, so that the posterior results in
\begin{equation}
    p(\bs{\phi} \mid \mathbf{y}, \mathbf{Z}, \bs{\delta}) \propto \mathcal{L}(\bs{\phi};\mathbf{y}, \mathbf{Z}, \bs{\delta}) p(\bs{\phi}) = \mathcal{L}(\bs{\phi};\mathbf{y}, \mathbf{Z}, \bs{\delta}) \prod_{k=1}^{3} p_{k}(\bs{\phi}_{k}),
    \label{eq:posterior}
\end{equation}
where $p_{k}(\bs{\phi}_{k})$ is the prior of the $k$-th parameter block. Moreover, to simplify the Metropolis step implementation, we map each parameter from its support to the real line and propose a new state from a Gaussian distribution in the mapped space. More details about the single transformations and how to modify the acceptance rejection ratio are given in Section~\ref{sec:app_alg} of the Supplementary Material. Finally, the adaptive strategy by \citet{miasojedow2013adaptive} is adopted to tune the variance of the proposal distribution. For details about the adaptive MCMC schemes, we refer to \citet{andrieu2008tutorial,roberts2001optimal,atchade2005adaptive}.

\subsection{Parallel tempering MCMC}\label{sec:tempering}

A preliminary study showed how the posterior distribution, under our model, has numerous well-separated modes. Indeed, starting from different initial states, the classical random-walk Metropolis-within-Gibbs became trapped, each time, in a different mode, failing to recover the whole posterior distribution. To deal with this issue we adopted a parallel tempering MCMC \citep{woodard2009conditions}, also known as Metropolis Coupled Markov Chain Monte Carlo \citep{altekar2004parallel}.

The algorithm tackles the multi-modality issue by performing -- in parallel -- several MCMC steps at various \emph{inverse temperatures}. The latter constitute a set of $L$ parameters $\beta_{l} \in (0,1]$, $l = 1,\dots,L$ such that $1=\beta_{0} > \beta_{1} > \dots > \beta_{L} > 0$. In particular, each of the MCMC algorithms running in parallel is designed to draw values from a modified version of the posterior, given by
\begin{equation*}
    p(\bs{\phi} \mid \mathbf{y}, \mathbf{Z}, \bs{\delta})^{\beta_{l}} \propto [\mathcal{L}(\bs{\phi};\mathbf{y}, \mathbf{Z}, \bs{\delta}) p(\bs{\phi}) ]^{\beta_{l}}.
\end{equation*}
In this way, as $\beta_{l} \rightarrow 0$, i.e., the temperature gets hotter, the multi-modal posterior flattens, 
and the corresponding chain better explores the posterior's support. On the other side, the coldest chain, i.e., the one with $\beta_0=1$,  coincides with the posterior of interest. Importantly, the states at various temperatures are continuously proposed to be switched during the run and a Metropolis step is employed to accept or reject the switch, so that information coming from the flattest chain ($\beta_{L}$) is allowed to reach the original one ($\beta_{0}=1$). Consequently, the coldest chain is facilitated in exploring all the modes of the posterior distribution. More precisely, at any iteration $g$, after the Metropolis-within-Gibbs step, for each block of the parameters, given the $L$ current values of the chains $\left[\bs{\phi}_{1}^{(g)},\dots,\bs{\phi}_{L}^{(g)}\right]$, the swap is proposed by uniformly sampling $\bar{l}$ from $\{1,\dots,L\}$, and swapping the current states at temperature levels $\bar{l}$ and $\bar{l} + 1$ with probability given by 
\begin{equation*}
    \omega\left(\bs{\phi}_{\bar{l}}^{(g)},\bs{\phi}_{\bar{l} + 1}^{(g)} \mid \mathbf{y}, \mathbf{Z}, \bs{\delta} \right) = \min \left\{ 1, \left( \frac{p\left( \bs{\phi}_{\bar{l} + 1}^{(g)} \mid \mathbf{y}, \mathbf{Z}, \bs{\delta} \right)}{p\left( \bs{\phi}_{\bar{l}}^{(g)} \mid \mathbf{y}, \mathbf{Z}, \bs{\delta} \right)} \right)^{\beta_{\bar{l}}-\beta_{\bar{l} + 1}} \right\}.
\end{equation*}

Clearly, when implementing a tempering MCMC, the choice of the inverse temperature parameters $\beta_l$, $l=1, \dots, L$ is crucial. Indeed, selecting temperatures that are too similar to  one another can lead to inadequate exploration of the parameter space. On the contrary, selecting temperatures that are too far apart from one another may lead to significantly low acceptance rates for swaps. To avoid this issues, following \cite{miasojedow2013adaptive}, we opt for an adaptive strategy to tune a function of the temperatures spacing, i.e., $\rho_{l} = \log ( 1/\beta_{l+1} - 1/\beta_{l} )$. Specifically, we fix the  number of temperature levels $L$ and target a specific mean acceptance rate of the swaps between the chains in nearby temperatures. Following the guidelines by \citep{atchade2005adaptive}, we set this value to $0.41$. More in details, at each iteration $g$ we update $\rho_{\bar{l}}$ by adding a function of the difference between the current accept/reject probability $\omega$ and the targeted one. In this way, if the swap acceptance probability is too high, the log-distance between $\bar{l}$ and $\bar{l} + 1$ temperatures increases, and, if it is too low, it decreases. 

We point  out that \citet{miasojedow2013adaptive} provide theoretical guarantees that the adaptive tempered algorithm eventually expresses the fixed acceptance rate for both the Metropolis acceptance rate within each block and the adjacent states swap. The same paper shows that the $L$ chains drawn in parallel by the algorithm are geometrically ergodic, so that both the strong law of large number and the central limit theorem hold true for these chains. A detailed description of the algorithm is provided in Section~\ref{sec:app_alg} of the Supplementary Material.

As a final note, it is important to consider that traditional methods to summarize multi-modal posterior densities, such as the posterior mean, median, or quantile-based credible intervals, may not be suitable. In this study, we use the posterior mode as point estimate of the parameter of interest, including the epicentre, while uncertainty is evaluated using highest posterior density regions (HPDRs). HPDRs are particularly useful in the case of multi-modal distributions, as they can include multiple disjoint subsets, one for each local mode, providing a more accurate representation of the underlying posterior framework.

\section{Simulation study}\label{sec:simulation}

We carry out a simulation study to show the soundness and the performance of our modelling approach under different smartphone network geometries. In this context, the network geometry is described in terms of the number of smartphones in the network, $n$, and in terms of their spatial variability in the north-south (N-S) and east-west (E-W) directions. 

Smartphone coordinates are randomly generated from a bivariate normal distribution with mean vector $(0,0)$ degrees and with diagonal covariance matrix $\sigma^2_s I_2$. Considering $n=\{25,50,100\}$ and $\sigma^2_s=\{1,0.25,0.05\}$, a total of 9 scenarios are defined. Under each scenario, $100$ simulation runs are executed. 
For each simulation run we randomly generate the smartphone coordinates $\mathbf{z}_i$, the earthquakes parameters (i.e., epicentre $\mathbf{z}_{0}$, depth $d_{0}$ and origin time $t_{0}$), the wave mixture parameter $\alpha$ and the  mixture cure parameter $\pi$. Then, $(100\pi)\%$ of the smartphones are randomly selected as cured. For the remaining smartphones, triggering times are randomly generated from the inverse of the mixture survival function. 

\begin{figure}[htbp!]
    \centering
    \includegraphics[width = 0.66\textwidth]{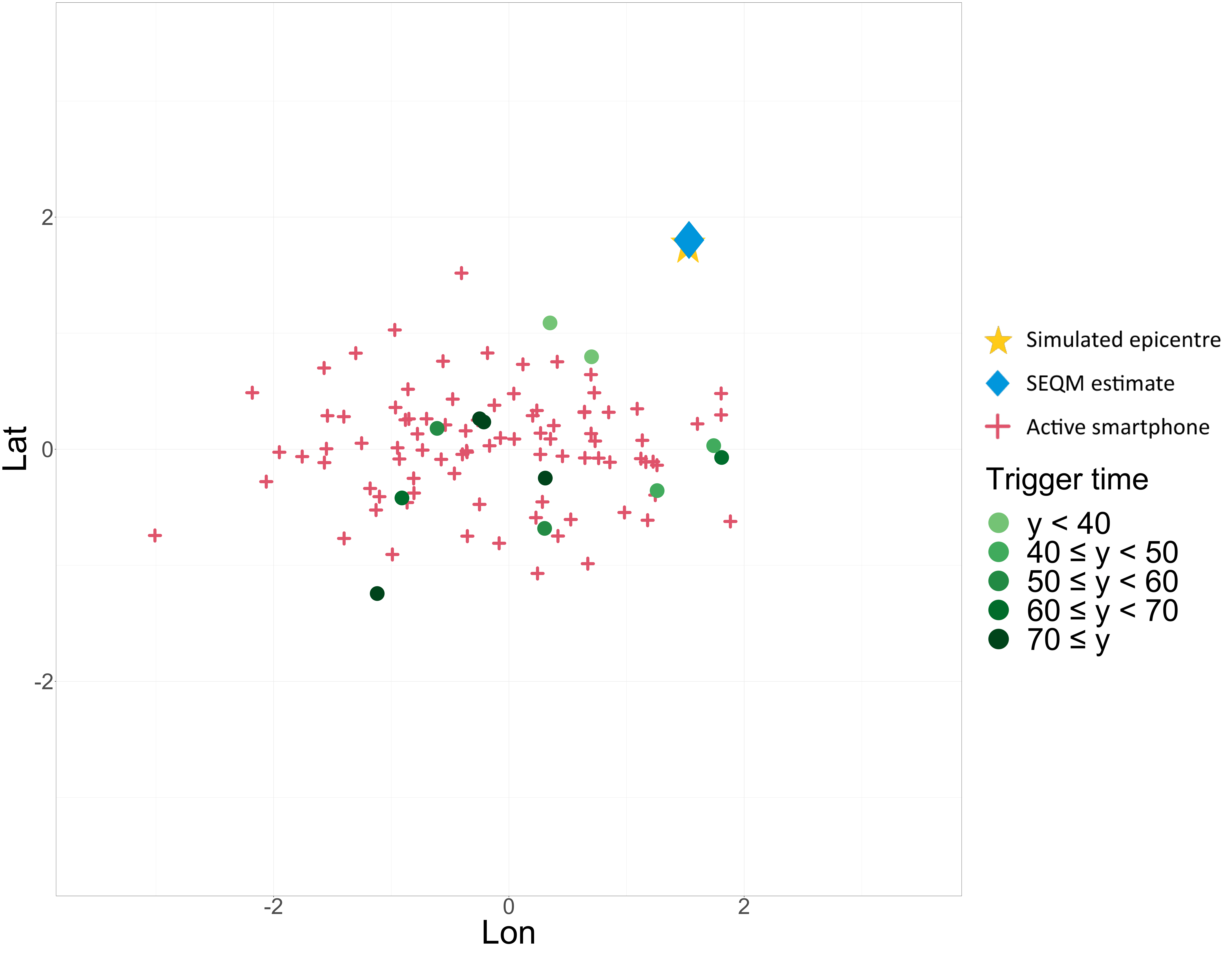}
    \caption{$n=100$ simulated data with network N-S and E-W spatial variances equal to $(0.25,1)$.}
    \label{fig:sim_map}
\end{figure}

In particular, model and earthquake parameters are generated from uniform distributions: epicentre $\mathbf{z}_{0}$ in the spatial domain $[-3,3]\times[-3,3]$ degrees, depth $d_{0}$ in the $[0,100]$ km interval, origin time $t_{0}$ in $[5,30]$ seconds, $\alpha$ in $[0,1]$, $\pi$ in $[0.5,0.95]$. Finally, the earthquake detection time $T^{*}$ is randomly generated from a uniform between 60 and 120 seconds after the true origin time. This determines which triggering times are actually observed and which are censored. An example of the simulated data along with the SEQM posterior estimate of the epicentre is given in Figure~\ref{fig:sim_map}.
In Section~\ref{sec:EQN}, we highlighted that the distribution of triggering times exhibits a high degree of variability and noise. This is evident also in Figure~\ref{fig:sim_map}, where the simulated data reflects this feature thanks 
the survival function outlined in Equation \eqref{eq:S}.

Posterior inference is carried out through the parallel tempering MCMC method described in Section~\ref{sec:comp_det}. The results for each simulation run are based on $25000$ iterations, where  the first $25000$ are discarded as burn-in period. For each run, point estimates of the earthquake parameters are obtained as the predominant mode of the corresponding posterior distributions.

\begin{figure}[htbp!]
    \centering
    \begin{tabular}{lll}
    (a) & (b) & (c) \\
        \includegraphics[width = 0.32\textwidth]{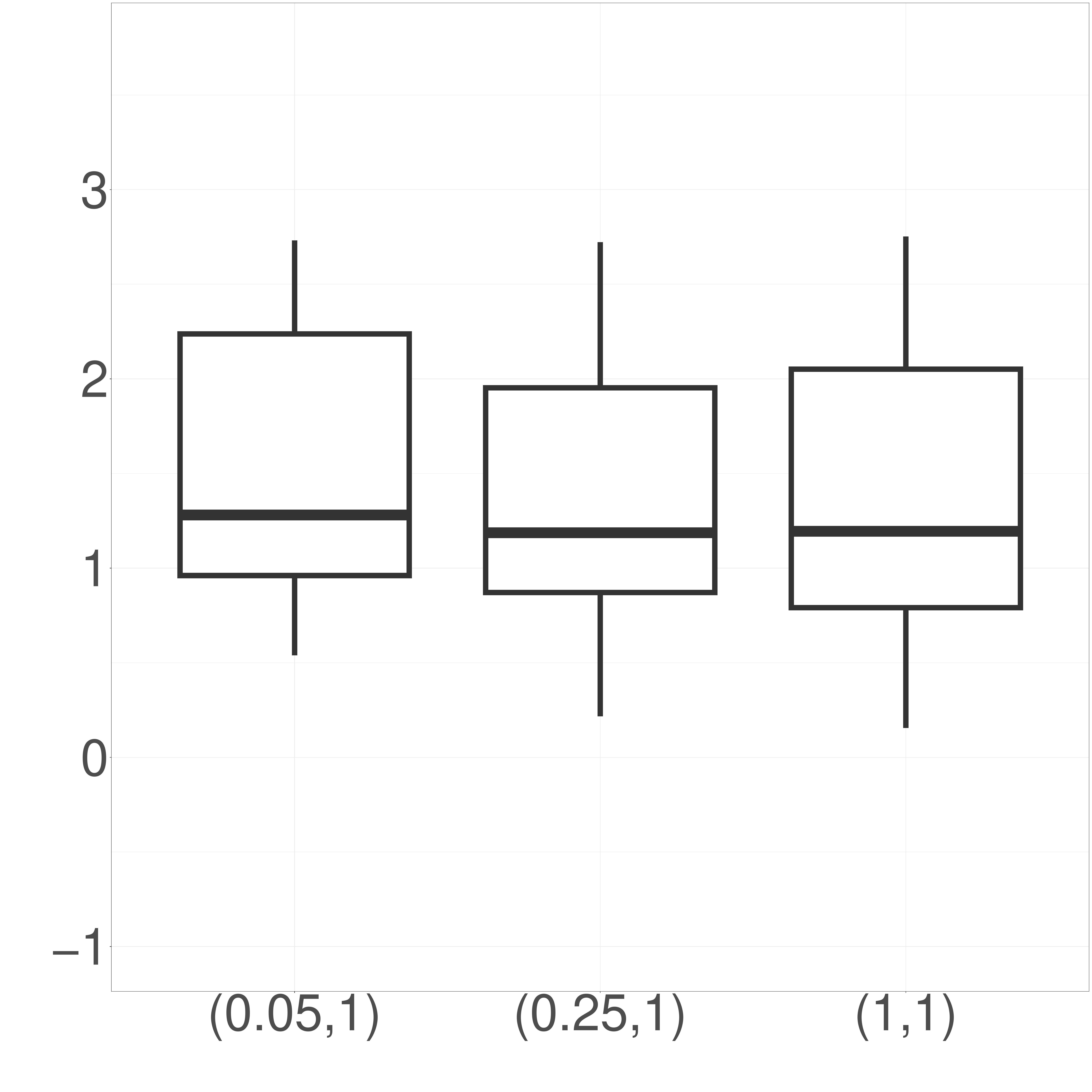}
    &
        \includegraphics[width = 0.32\textwidth]{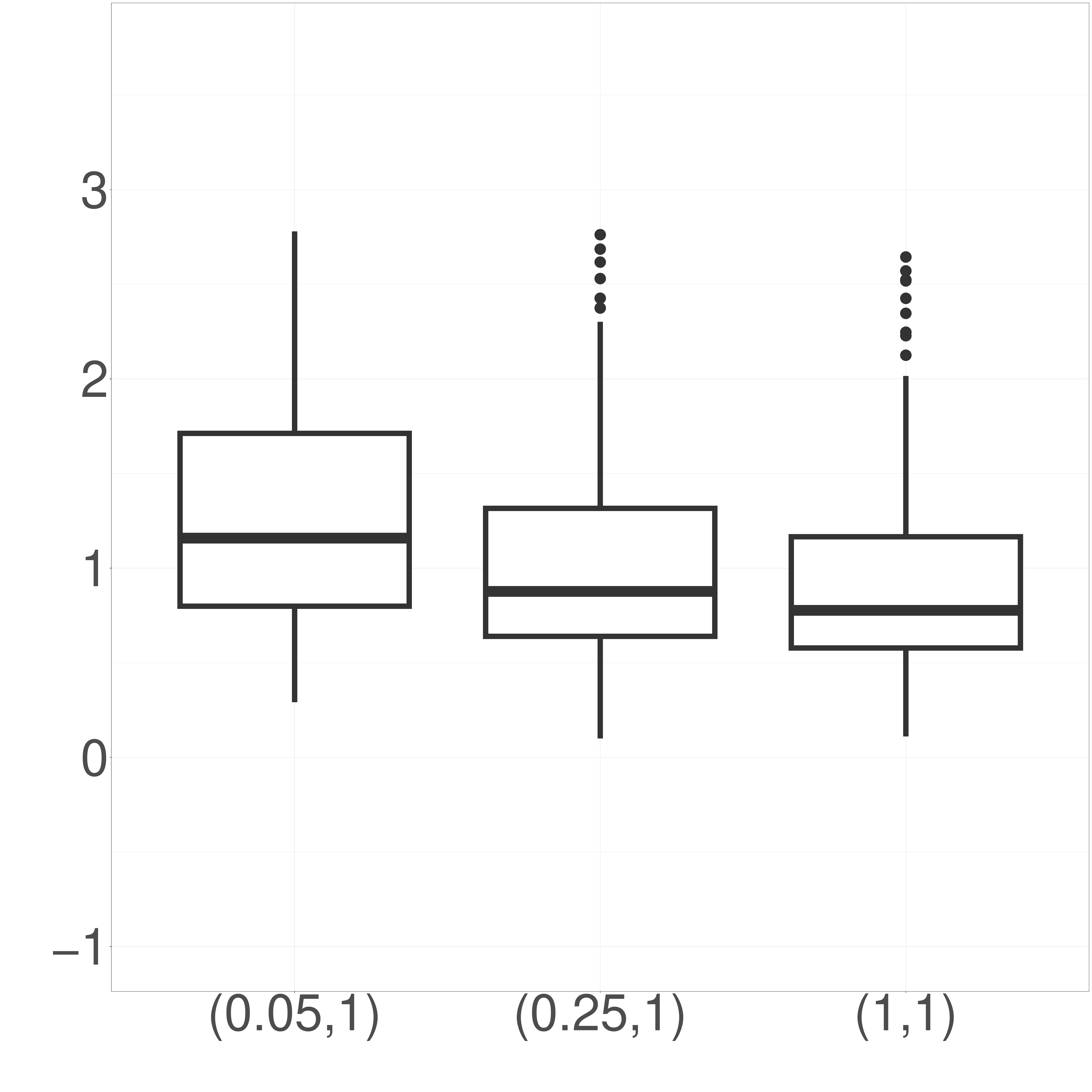}
    &
        \includegraphics[width = 0.32\textwidth]{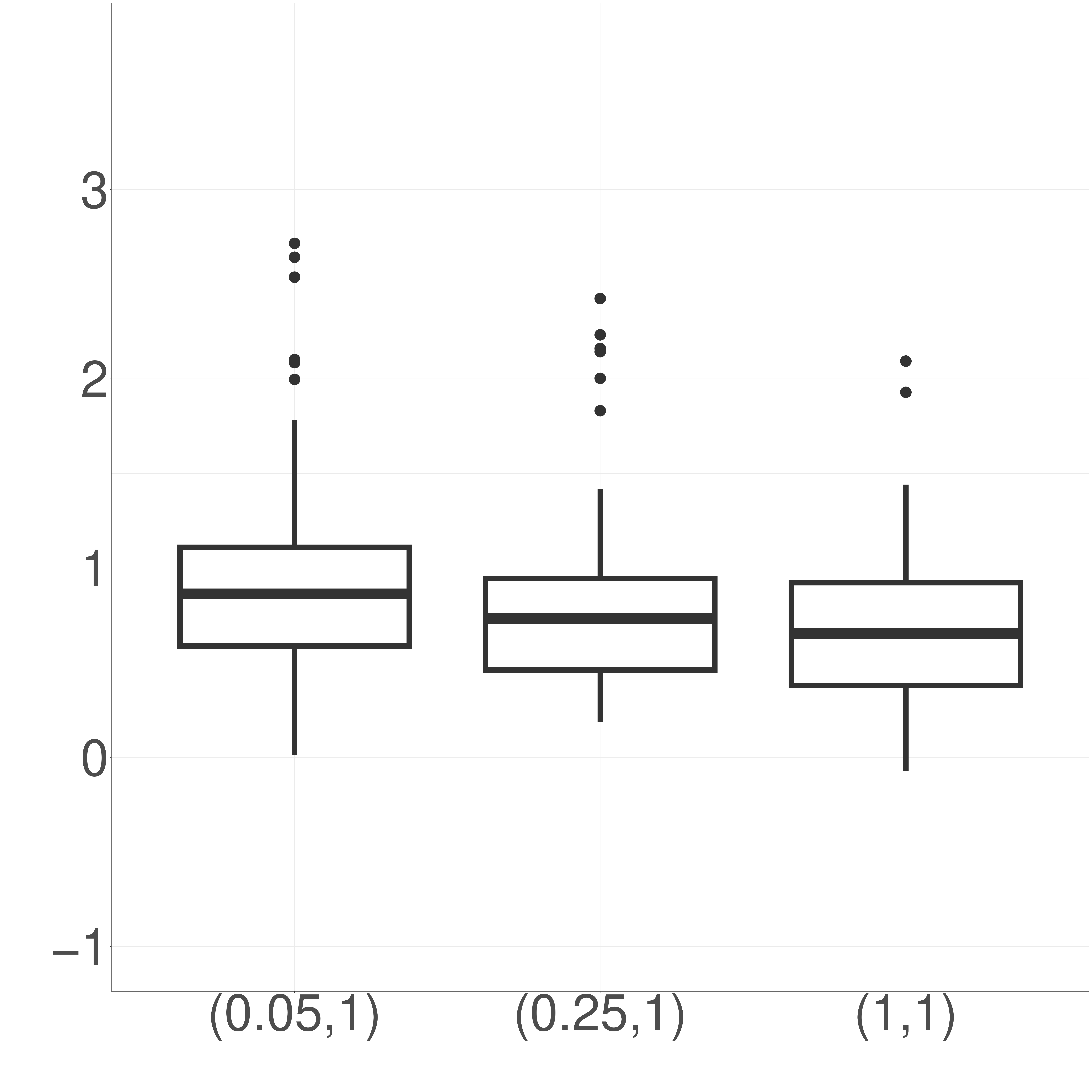}
    \end{tabular}
    \caption{Box-plots for epicentre error $n=25$ (a), $n=50$ (b) and $n=100$ (c). On the y-axis it is represented the $\log_{10}$ of the SEQM epicentre estimation error while on the x-axis the network N-S and E-W spatial variances.}
    \label{fig:sim_boxplots}
\end{figure}

\begin{figure}[htbp!]
    \centering
    \begin{tabular}{lll}
    (a) & (b) & (c) \\
        \includegraphics[width = 0.32\textwidth]{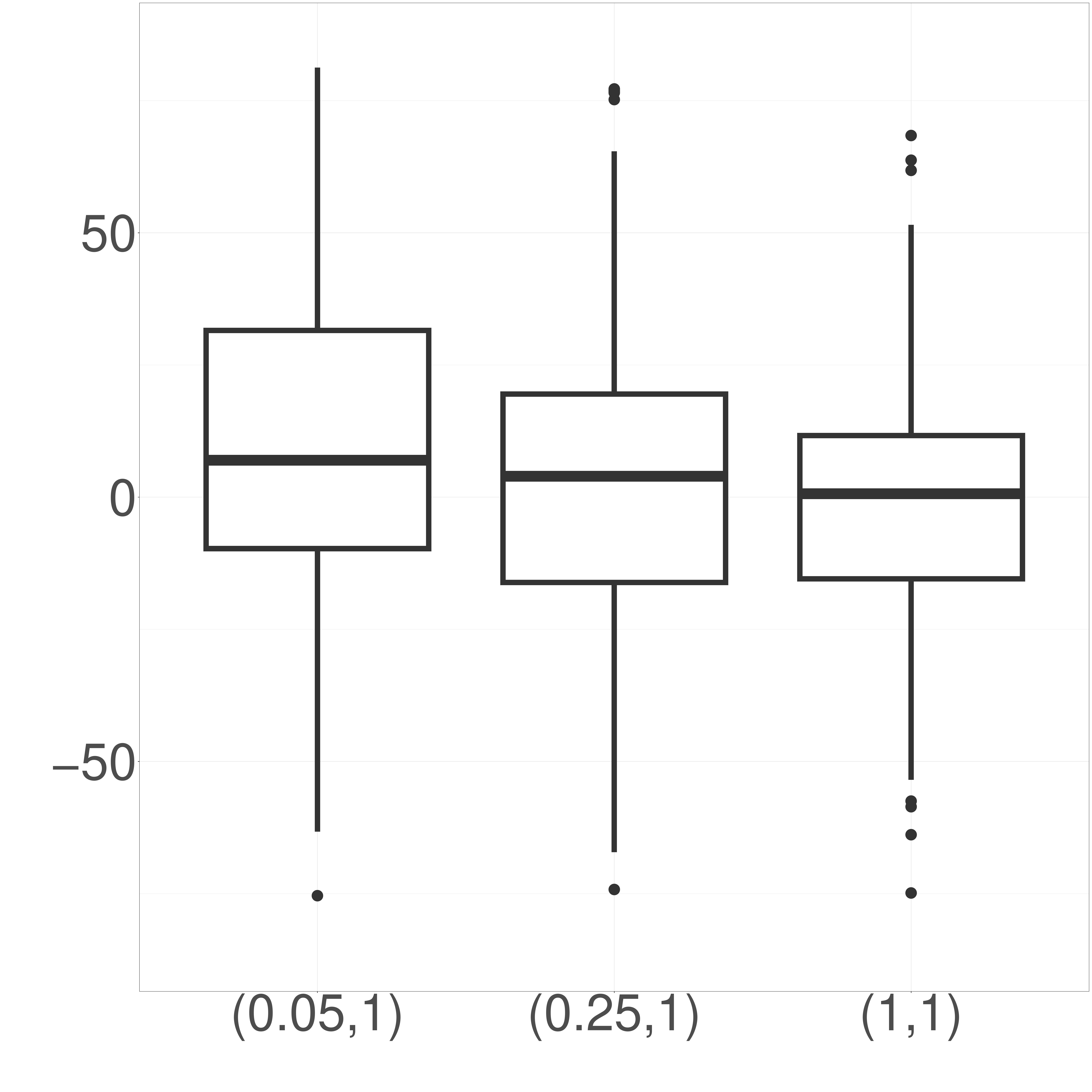}
    &
        \includegraphics[width = 0.32\textwidth]{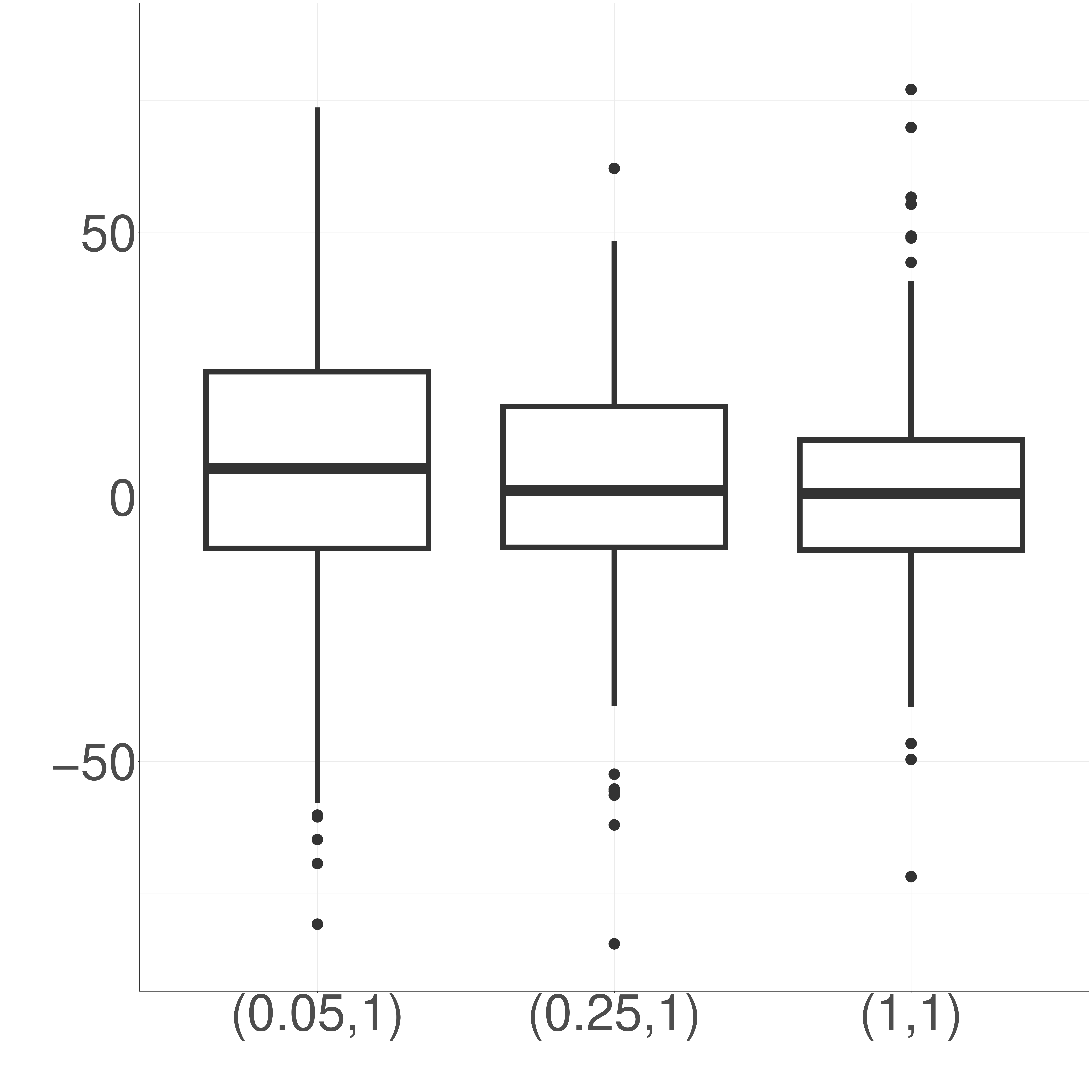}
    &
        \includegraphics[width = 0.32\textwidth]{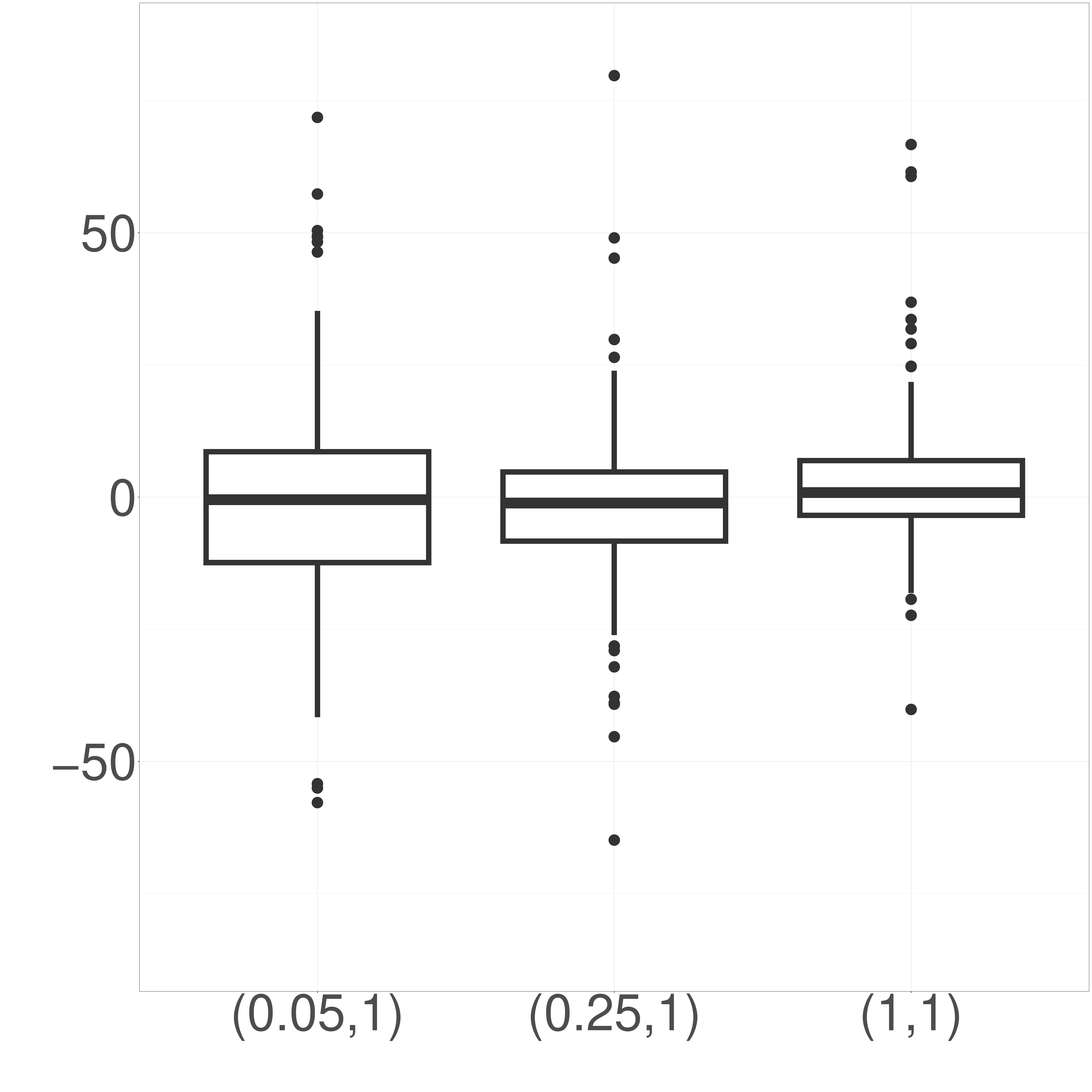}
    \end{tabular}
    \caption{Box-plots for origin time error $n=25$ (a), $n=50$ (b) and $n=100$ (c). On the y-axis it is represented the SEQM origin time estimation error while on the x-axis the network N-S and E-W spatial variances.}
    \label{fig:sim_t_boxplots}
\end{figure}

Box-plots in Figure~\ref{fig:sim_boxplots} describe the distribution of the geodetic distance (error) between estimated and simulated epicentre under the different scenarios. As expected, both the median error and the variability of the error tend to decrease when $n$ and/or $\sigma^2_s$ increase. Analogous considerations can be made for the results in Figure~\ref{fig:sim_t_boxplots}, which shows the distributions of the origin time error, i.e., the difference between the simulated origin time and the SEQM estimated time.

\section{Analysis of EQN datasets}\label{sec:case_study}

\begin{figure}
    \centering
         \includegraphics[width = 0.49\textwidth]{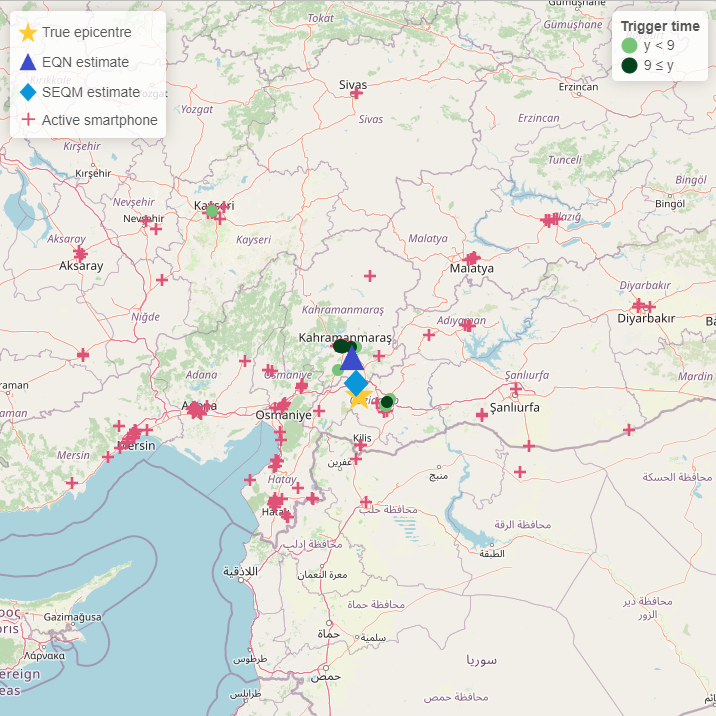}
        \includegraphics[width = 0.49\textwidth]{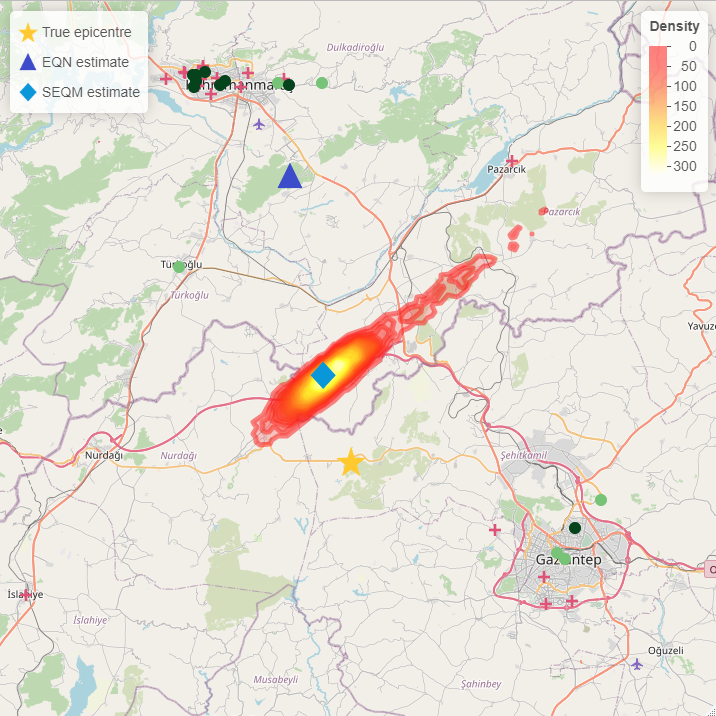}
    \caption{Summary map (left panel) and posterior density of the epicentre location (right panel) for the Turkish-Syrian earthquake. The zero of the triggering times is the earthquake origin time.
    }
    \label{fig:summary_turkey}
\end{figure}

\begin{figure}
    \centering
        \includegraphics[width = 0.49\textwidth]{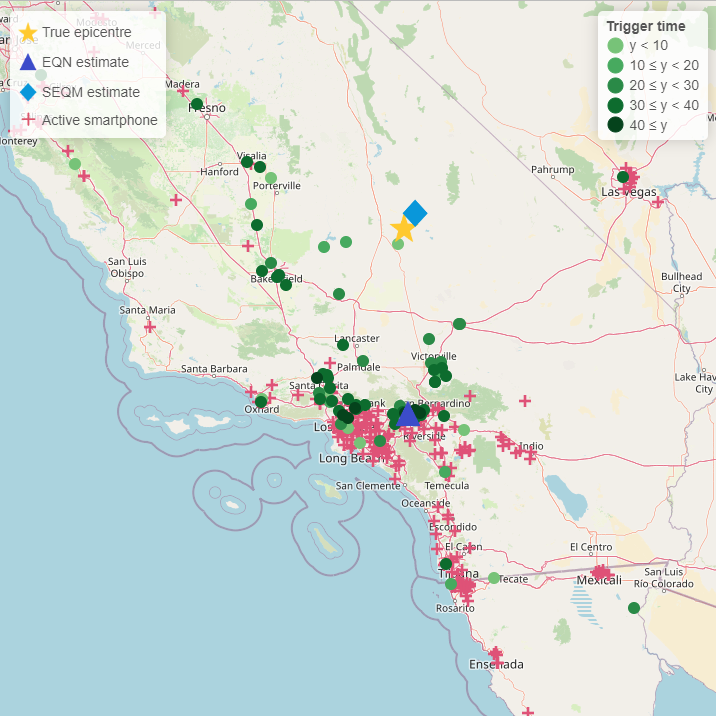}
        \includegraphics[width = 0.49\textwidth]{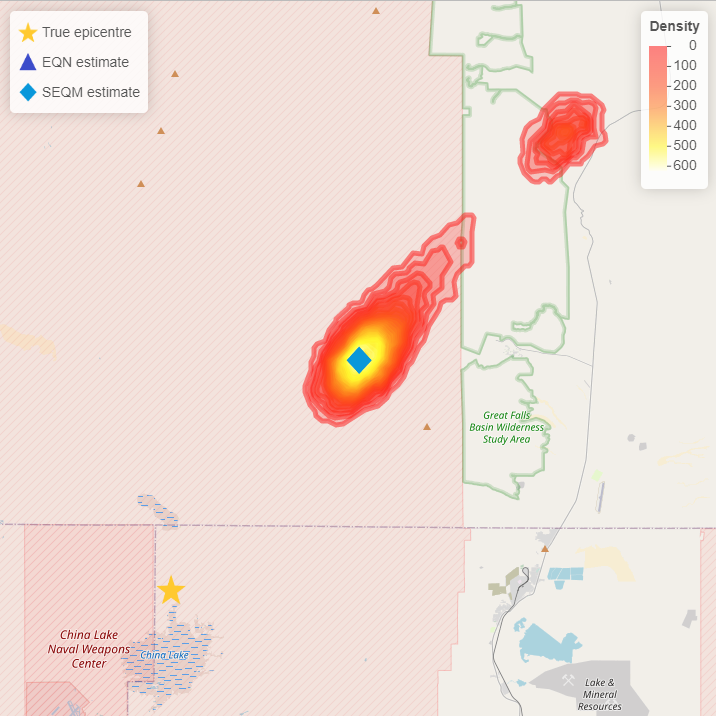}
    \caption{Summary map (left panel) and posterior density of the epicentre location (right panel) for the Ridgecrest earthquake. The zero of the triggering times is the earthquake origin time. 
    }
    \label{fig:summary_california}
\end{figure}

\begin{figure}
    \centering
        \includegraphics[width = 0.49\textwidth]{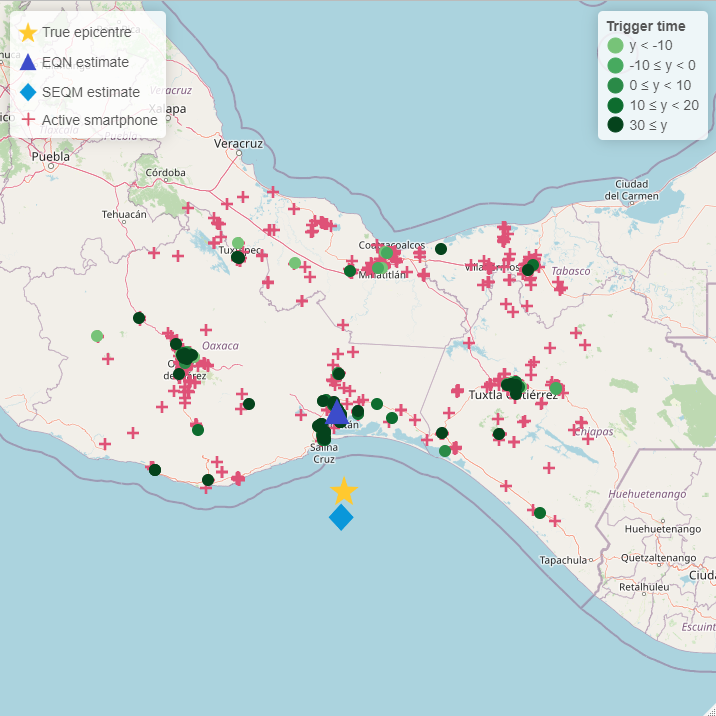}
         \includegraphics[width = 0.49\textwidth]{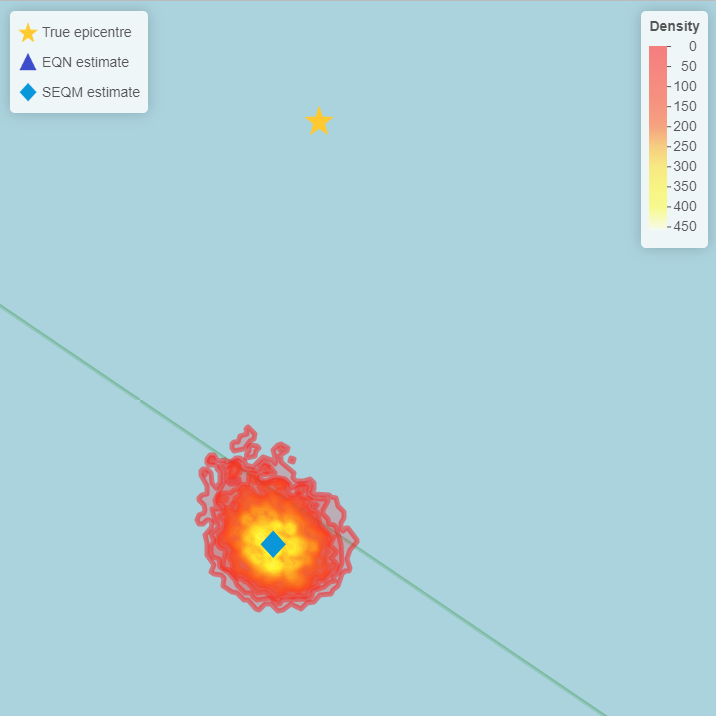}
    \caption{Summary map (left panel) and posterior density of the epicentre location (right panel) for the Mexican earthquake. The zero of the triggering times is the earthquake origin time. 
    }
    \label{fig:summary_mexico}
\end{figure}

For each dataset described in Section \ref{sec:EQN_datasets}, earthquake parameters are estimated employing the survival model outlined in Equation \eqref{eq:likelihood}, with the prior defined in Section~\ref{sec:prior} and using the algorithm detailed in Section~\ref{sec:comp_det}.  
The posterior summaries of model and earthquake parameters are provided in Section~\ref{sec:app_images} of the Supplementary Material.
Here, we report
Table~\ref{tab:summary_table} that includes the posterior mode and the associated $95\%$ HPDRs of the earthquake parameters and the $\alpha$ and $\pi$ model parameters for the three case studies. The table also reports the EQN estimates of the earthquake parameters. While EQN only provides point estimates, SEQM also offers uncertainty quantification through the HPDRs. 
The summary maps in  
Figures \ref{fig:summary_turkey} - \ref{fig:summary_mexico}  show the epicentre estimates provided by EQN and SEQM (left panels), together with the true epicentre of the three earthquakes. The right panel of the figures displays the posterior density of the epicentre location. 

In Table~\ref{tab:comparison} we the assess accuracy by measuring the distance between the true and the estimated epicentre, between the true and the estimated depth (only provided by SEQM), and between the true and the estimated origin time. Overall, SEQM is characterised by a lower error thus improving EQN estimates.

\begin{table}
    \caption{\label{tab:summary_table} EQN and SEQM earthquake parameter estimates for the three events. For SEQM, the $95\%$ HPDR subset corresponding to the predominant mode is also provided.}
    \centering
    \begin{tabular}{l|lrrr}
    &  & Turkey-Syria & Ridgecrest & Mexico \\ 
    \multirow{3}{*}{EQN} & $lat^{\ast }$ [deg] & \multicolumn{1}{r}{37.48} & 
    \multicolumn{1}{r}{34.08} & \multicolumn{1}{r}{16.47} \\ 
    & $lon^{\ast }$ [deg] & \multicolumn{1}{r}{37.00} & \multicolumn{1}{r}{-117.57}
    & \multicolumn{1}{r}{-95.05} \\ 
    & $t^{\ast }$ [hh:mm:ss] & \multicolumn{1}{r}{01:17:46} & \multicolumn{1}{r}{
    03:20:33} & \multicolumn{1}{r}{06:26:17} \\ \hline
    \multirow{12}{*}{SEQM} & $lat_{0}$ [deg] & \multicolumn{1}{r}{37.26} & 
    \multicolumn{1}{r}{35.89} & \multicolumn{1}{r}{15.36} \\ 
    &  & \multicolumn{1}{r}{ [37.19, 37.43]} & \multicolumn{1}{r}{ %
    [35.86, 35.95]} & \multicolumn{1}{r}{ [15.32, 15.41]} \\ 
    & $lon_{0}$ [deg] & \multicolumn{1}{r}{37.04} & \multicolumn{1}{r}{-117.49}
    & \multicolumn{1}{r}{-95.01} \\ 
    &  & \multicolumn{1}{r}{ [36.94, 37.28]} & \multicolumn{1}{r}{ %
    [-117.53, -117.43]} & \multicolumn{1}{r}{[-95.05, -94.96]} \\
    & $d_{0}$ [km] & \multicolumn{1}{r}{9.97} & \multicolumn{1}{r}{92.03} & 
    \multicolumn{1}{r}{15.86} \\ 
    &  & \multicolumn{1}{r}{[3.54, 45.45]} & \multicolumn{1}{r}{%
    [80.89, 99.75]} & \multicolumn{1}{r}{[4.42, 40.74]} \\ 
    & $t_{0}$ [hh:mm:ss] & \multicolumn{1}{r}{01:17:34} & \multicolumn{1}{r}{
    03:19:40} & \multicolumn{1}{r}{06:25:46} \\ 
    &  & \multicolumn{1}{r}{[01:17:29, 01:17:35]} & \multicolumn{1}{r}%
    {[03:19:37, 03:19:41]} & \multicolumn{1}{r}{[06:25:46,
    06:25:47]} \\
      & $\alpha$ & \multicolumn{1}{r}{0.00} & \multicolumn{1}{r}{0.02} & \multicolumn{1}{r}{0.34} \\ 
      & & \multicolumn{1}{r}{[0.00, 0.24]} & \multicolumn{1}{r}{[0.00, 0.05]} & \multicolumn{1}{r}{[0.24, 0.48]} \\
       & $\pi$ & \multicolumn{1}{r}{0.68} & \multicolumn{1}{r}{0.46} & \multicolumn{1}{r}{0.81} \\ 
    && \multicolumn{1}{r}{[0.50, 0.82]} & \multicolumn{1}{r}{[0.34, 0.56]} & \multicolumn{1}{r}{[0.77, 0.86]}
    \end{tabular}
\end{table}

\begin{table}
    \caption{\label{tab:comparison} Comparison between EQN and SEQM earthquake parameter estimates. Errors are computed with respect to the true earthquake parameter in Table \ref{tab:cs_details}.}
    \centering
    \begin{tabular}{l|lrrr}
    &  & Turkey-Syria & Ridgecrest & Mexico \\ \hline
    \multirow{2}{*}{EQN error} & epicentre [km] & \multicolumn{1}{r}{35.40} & 
    \multicolumn{1}{r}{187.43} & \multicolumn{1}{r}{92.86} \\ 
    & origin time [s] & \multicolumn{1}{r}{10.00} & \multicolumn{1}{r}{41.00} & 
    \multicolumn{1}{r}{29.00} \\ \hline
    \multirow{3}{*}{SEQM error} & epicentre [km] & \multicolumn{1}{r}{11.02} & 
    \multicolumn{1}{r}{18.34} & \multicolumn{1}{r}{31.39} \\ 
    & depth [km] & \multicolumn{1}{r}{10.03} & \multicolumn{1}{r}{84.03} & 
    \multicolumn{1}{r}{12.01} \\ 
    & origin time [s] & \multicolumn{1}{r}{1.86} & \multicolumn{1}{r}{11.80} & 
    \multicolumn{1}{r}{1.37}%
    \end{tabular}
\end{table}

In particular, for the Turkish-Syrian event, epicentre and depth are estimated with an error of around 11 and 10 km, respectively, while the error on the origin time is only around 2 seconds. The estimated $\alpha$ points out that nearly all smartphones triggered on the S wave, while the estimated $\pi$ indicates that three-quarters of the smartphones did not trigger even if the seismic waves reached them (\emph{cured} smartphones).

For the Ridgecrest event, the epicentre estimation error is around 18 km, which is notable considering that the EQN detection occurred at high distance. The error on the depth, however, is around 84 km while the estimated origin time is around 12 seconds earlier that the true time. This likely indicates that the P and/or S velocities used in the SEQM are higher than those actually observed. The estimated $\alpha$ and $\pi$ suggest that the smartphones primarily detected the S wave and that half of them were cured.

For the Mexican event, the error on the epicentre location is around 31 km, smaller than the 93 km error given by EQN. Error on depth is around 12 km while the error on origin time is around 1.4 seconds. The estimated $\alpha$ indicates that 1/3 of the devices detected the P wave, while $\pi$ indicates that more than $80\%$ of the devices were cured.

\section{Discussion and conclusions}\label{sec:discussion}

Smartphone networks employed in earthquake early warning are complex stochastic objects characterised by a random geometry and by nodes (the smartphones) which exhibit a non-predictable behaviour during an earthquake. In this paper we developed a modelling approach for estimating relevant earthquake parameters, uncertainty included, starting from the data provided by the smartphone network when an earthquake is detected in real-time. The approach, which is largely inspired by survival analysis, allows to handle the uncertainty on the smartphone data and to fully exploit the information content of the dataset. More precisely, earthquake parameter estimation is not only affected by the trigger spatial locations and times but also by the spatial locations of the non-triggering smartphones. This helps to constraint the epicentre location and origin time, especially when the number of triggers is relatively small and/or all the triggers are clustered within a small area.

A simulation study showed that the model capability to provide good estimates of the earthquake parameters is largely influenced by the smartphone network geometry and, more obviously, by the number of smartphones in the network. Better estimates are obtained when the network has a large spatial variability along the north-south and east-west directions. This induces a larger variability on the triggering times, which, again, is needed to constraint the earthquake parameters.

When applied to real EQN datasets, our modelling approach largely improved the EQN estimates of epicentre and origin time, while providing measures of uncertainty. Posterior densities of the epicentre location, however, exhibit a small spatial variability and they do not encompass the true epicentre. This is possibly a consequence of the velocity model adopted in Equation \eqref{eq:gauss_mixt_par}, with the seismic waves having fixed deterministic velocities. This problem may be likely mitigated describing the P and S wave velocities as random variables by introducing prior distributions on them.

In addition, future developments may include implementing a dependent data prior approach. We defined a diffuse prior on the hypocentre location and its estimate is strongly informed by the smartphone data through the likelihood. However, hypocentres are more likely to be located near known faults. Earthquake catalogues may thus be exploited to define more informative priors to better constraint estimates.

Finally, optimising the computational efficiency for real-time usage is a key developments that will be addressed.

\section*{Funding}

This article was partially funded by the European
Union (EU)’s Horizon 2020 Research and Innovation Program under Grant Agreement RISE Number 821115. Opinions expressed in this article solely reflect the authors’ views; the EU is not responsible for any use that may be made of information it contains.

\section*{Supplementary Material}

\renewcommand\thefigure{S\arabic{figure}}
\setcounter{figure}{0}    
\begin{appendix}
    \section{MCMC scheme}\label{sec:app_alg}
    Recall that in Section~\ref*{sec:comp_det} we have defined $\bm{\phi} = (\bm{\phi}_{1},\bm{\phi}_{2},\bm{\phi}_{3}) =(\bm{\theta},\alpha,\pi)$. From Equation~(\ref*{eq:posterior}), we define $p_{k}(\bm{\phi}_{k} \mid \mathbf{y}, \mathbf{Z}, \bm{\delta}) \propto \mathcal{L}(\bm{\phi}; \mathbf{y}, \mathbf{Z}, \bm{\delta}) p_{k}(\bm{\phi}_{k})$ the full conditional distribution of each block of $\bm{\phi}$. Let $J_k\left(\widetilde{\bm{\phi}}_{k}\right)$ be the Jacobian of the transformation function $m_{k}: \mathbb{R}^{d_{k}} \rightarrow S_{k}$, where $S_{k} = [lb_{1_{k}},ub_{1_{k}}] \times \dots \times [lb_{d_{k}},ub_{d_{k}}]$, $lb_{i_{k}}$ and $ub_{i_{k}}$ are the bounds of the support of the $i_{k}$-th parameter within the $k$-th block, $d_{k}$ is the dimension of the $k$-th block and $\widetilde{\bm{\phi}}_{k} = m_{k}^{-1}\left(\bm{\phi}_{k}\right)$.
    We generalise here the notation by referring to $\bm{\phi}_{k,l}^{(g)}$ as the value at iteration $g$ of the $k$-th sub-vector of $\bm{\phi}_{l}^{(g)} = \left(\bm{\phi}_{1,l}^{(g)},\dots,\bm{\phi}_{K,l}^{(g)}\right)$ regarding the chain targeting the $l$-th version of the posterior, i.e., $p\left(\bm{\phi}_{l} \mid \mathbf{y}, \mathbf{Z}, \bm{\delta}\right)^{\beta_{l}}$, with $l=1,\dots,L$.
    
    Algorithm \ref{alg:atmwg} provides the details of the MCMC scheme we designed to approximate the joint posterior distribution under the SEQM. 
    In our application, we assumed $K = 3$ blocks with dimension $\mathbf{d} = (d_{1},d_{2},d_{3}) = (4,1,1)$  and  $L=10$. 
    Regarding the transformations, the parameter supports are $[lb_{1_{1}},ub_{1_{1}}] = [-90,90]$, $[lb_{2_{1}},ub_{2_{1}}] = [-180,180]$, $[lb_{3_{1}},ub_{3_{1}}] = [5,100]$, $[lb_{4_{1}},ub_{4_{1}}] = [0,\infty)$, $[lb_{1_{2}},ub_{1_{2}}] = [0,1]$ and $[lb_{1_{3}},ub_{1_{3}}] = [0,1]$. Hence, defining $\widetilde{\bm{\phi}}_{1} = \left(\widetilde{lat}_{0},\widetilde{lon}_{0},\tilde{d}_{0},\tilde{t}_{0}\right)$, $\widetilde{\bm{\phi}}_{2} = \widetilde{\alpha}$ and $\widetilde{\bm{\phi}}_{3} = \widetilde{\pi}$, transformation functions are defined as:
    \begin{align*}
        m_{1}\left(\widetilde{\bm{\phi}}_{1}\right) &= \left(g\left(\widetilde{lat}_{0};-90,90\right),g\left(\widetilde{lon}_{0};-180,180\right),g\left(\tilde{d}_{0};5,100\right),e^{\tilde{t}_{0}}\right) \\
        m_{2}\left(\widetilde{\bm{\phi}}_{2}\right) &= g\left(\widetilde{\alpha};0,1\right) \\
        m_{3}\left(\widetilde{\bm{\phi}}_{3}\right) &= g\left(\widetilde{\pi};0,1\right)
    \end{align*}
    where 
    \begin{equation*}
        g(\widetilde{x};lb,ub)= \frac{lb + ub \, e^{\widetilde{x}}}{1 + e^{\widetilde{x}}}
    \end{equation*}
    
    For each block and for $l=1,\dots,10$ we set the starting values of the model parameters as $\widetilde{lat}_{0}^{(0)} = g^{-1}\left(lat_{0}^{(0)};-90,90\right)$, $\widetilde{lon}_{0}^{(0)} = g^{-1}\left(lon_{0}^{(0)};-180,180\right)$, $\tilde{d}_{0}^{(0)} = g^{-1}\left(d_{0}^{(0)};5,100\right)$, $\tilde{t}_{0}^{(0)} = \log{\left(t_{0}^{(0)}\right)}$ and set $\widetilde{\bm{\phi}}_{1,l}^{(0)} = \left(\widetilde{lat}_{0}^{(0)},\widetilde{lon}_{0}^{(0)}, \tilde{d}_{0}^{(0)},\tilde{t}_{0}^{(0)}\right)$. In particular $\mathbf{z}_{0}^{(0)}=\left(lat_{0}^{(0)},lon_{0}^{(0)}\right)$ is sampled from a uniform centered on $\mathbf{z}^{\ast}$ with a two degree wide support, $d_{0}^{(0)}$ is sampled from a uniform between $5$ and $100$ km and $t_{0}^{(0)}$ is sampled from a uniform centered on the EQN origin time detection, with a $20$ seconds wide support. For $\widetilde{\bm{\phi}}_{2,l}^{(0)} = g^{-1}\left(\alpha_{0}^{(0)};0,1\right)$ and $\widetilde{\bm{\phi}}_{3,l}^{(0)} = g^{-1}\left(\pi_{0}^{(0)};0,1\right)$ we sample each of them from a uniform distribution between $0$ and $1$.
    Moreover, we fix the target acceptance rates $\Bar{\bm{\xi}} = \left(\Bar{\xi}_{1},\Bar{\xi}_{2},\Bar{\xi}_{3}\right) = (0.23,0.41,0.41)$, the adaptation stepsize sequences $\nu = 0.6$, and the target adjacent state swapping rate $\Bar{\omega} = 0.41$.
    Finally, regarding the initial values of the adaptation parameters, we set 
    chain means $\bm{\mu}_{k,l}^{(0)} = \widetilde{\bm{\phi}}_{k,l}^{(0)}$, scaling parameters $s_{k,l}^{(0)} = 0.1$, covariance estimates $\mathbf{R}_{1,l}^{(0)}= \diag(0.1,0.1,10,1)$, $\mathbf{R}_{2,l}^{(0)} = 0.1$, $\mathbf{R}_{3,l}^{(0)} = 0.1$, proposal covariances $\bm{\zeta}_{k,l}^{(0)} = \exp{\left(s_{k,l}^{(0)}\right)} \mathbf{R}_{k,l}^{(0)}$, and the temperatures spacing logarithm $\rho_{l}^{(0)} = 1$.

    \begin{algorithm}[htbp!]
        \caption{Adaptive Tempered Metropolis within Gibbs algorithm}\label{alg:atmwg}
        \begin{algorithmic}[1]
            \small
            \State Initialisation: the number of blocks $K$, their dimensions $\mathbf{d}$, the number of temperatures $L$, the required MCMC sample size $G$, burn-in period $g_{0}$, the starting values $\widetilde{\bm{\phi}}_{k,l}^{(0)}$, $\bm{\mu}_{k,l}^{(0)}$, $\bm{\zeta}_{k,l}^{(0)}$, $s_{k,l}^{(0)}$, $\mathbf{R}_{k,l}^{(0)}$ and $\rho_{l}^{(0)}$ for $k=1,\dots,K$, $l=1,\dots,L$, the acceptance probability rates $\Bar{\bm{\xi}}$, the adaptation stepsize sequences $\nu$, and the adjacent state swapping rate $\bar{\omega}$.
            \For{$g=1,2,\dots,G$}
                \For{$k=1,\dots,K$}
                    \For{$l=1,\dots,L$}
                        \State Sample $\widetilde{\bm{\phi}}_{k,l}^{\prime} = \widetilde{\bm{\phi}}_{k,l}^{(g-1)} + \bm{\epsilon}_{g,k,l}$ where $\bm{\epsilon}_{g,k,l} \sim \mathcal{N}_{d_{k}}\left(\mathbf{0},\bm{\zeta}_{k,l}^{(g-1)}\right)$
                        \State Let the next iterate be
                        \begin{equation*}
                            \widetilde{\bm{\phi}}_{k,l}^{(g)} = 
                            \begin{cases}
                                \widetilde{\bm{\phi}}_{k,l}^{\prime} &\text{with probability } \xi_{k,l}^{(g)}\left(\bm{\phi}_{k,l}^{(g-1)},\widetilde{\bm{\phi}}_{k,l}^{(g-1)},\bm{\phi}_{k,l}^{\prime},\widetilde{\bm{\phi}}_{k,l}^{\prime}\mid \mathbf{y}, \mathbf{Z}, \bm{\delta},\beta_{l}^{(g-1)}\right) \\[0.5em]
                                \widetilde{\bm{\phi}}_{k,l}^{(g-1)} &\text{with probability } 1 - \xi_{k,l}^{(g)}\left(\bm{\phi}_{k,l}^{(g-1)},\widetilde{\bm{\phi}}_{k,l}^{(g-1)},\bm{\phi}_{k,l}^{\prime},\widetilde{\bm{\phi}}_{k,l}^{\prime}\mid \mathbf{y}, \mathbf{Z}, \bm{\delta},\beta_{l}^{(g-1)}\right)
                           \end{cases}
                        \end{equation*}  
                         where
                         \begin{align*}
                             \bm{\phi}_{k,l}^{\prime} &= m_{k}\left(\widetilde{\bm{\phi}}_{k,l}^{\prime}\right) \\
                             \xi_{k,l}^{(g)}\left(\bm{\phi}_{k,l}^{(g-1)},\widetilde{\bm{\phi}}_{k,l}^{(g-1)},\bm{\phi}_{k,l}^{\prime},\widetilde{\bm{\phi}}_{k,l}^{\prime}\mid \mathbf{y}, \mathbf{Z}, \bm{\delta},\beta_{l}^{(g-1)}\right) &= \min\left\{1,\left(\frac{\
                            p_{k}\left(\bm{\phi}_{k,l}^{\prime}\mid \mathbf{y}, \mathbf{Z}, \bm{\delta}\right) \lvert J_k\left(\widetilde{\bm{\phi}}_{k,l}^{\prime}\right) \rvert }{p_{k}\left(\bm{\phi}_{k,l}^{(g-1)}\mid \mathbf{y}, \mathbf{Z}, \bm{\delta}\right) \lvert J_k\left(\widetilde{\bm{\phi}}_{k,l}^{(g-1)}\right) \rvert }\right)^{\beta_{l}^{(g-1)}}\right\}
                        \end{align*}
                        \State Compute
                        \begin{align*}
                            s_{k,l}^{(g)} &= s_{k,l}^{(g-1)}  + (g+1)^{-\nu} \left(\xi_{k,l}^{(g)}\left(\bm{\phi}_{k,l}^{(g-1)},\widetilde{\bm{\phi}}_{k,l}^{(g-1)},\bm{\phi}_{k,l}^{\prime},\widetilde{\bm{\phi}}_{k,l}^{\prime}\mid \mathbf{y}, \mathbf{Z}, \bm{\delta},\beta_{l}^{(g-1)}\right) - \Bar{\xi}_{k}\right) \\[0.5em]
                            \mathbf{R}_{k.l}^{(g)} &= \left(1-(g+1)^{-\nu}\right)\mathbf{R}_{k,l}^{(g-1)} + (g+1)^{-\nu} \left( \widetilde{\bm{\phi}}_{k,l}^{(g)} - \bm{\mu}_{k,l}^{(g-1)} \right) \left( \widetilde{\bm{\phi}}_{k,l}^{(g)} - \bm{\mu}_{k,l}^{(g-1)} \right)^\top \\[0.5em]
                            \bm{\mu}_{k,l}^{(g)} &= \left(1-(g+1)^{-\nu}\right)\bm{\mu}_{k,l}^{(g-1)} + (g+1)^{-\nu}\widetilde{\bm{\phi}}_{k,l}^{(g)}
                        \end{align*}
                        \State and
                        \begin{equation*}
                            \bm{\zeta}_{k,l}^{(g)} = \exp{\left(s_{k,l}^{(g)}\right)} \mathbf{R}_{k,l}^{(g)}
                        \end{equation*}
                    \EndFor
                \EndFor
                \State Sample $\bar{l} \in \{1,\dots,L-1\}$ uniformly and swap states $\bar{l}$ and $\bar{l}+1$ with probability
                \begin{equation*}
                    \omega_{\bar{l}}^{(g)}\left(\bm{\phi}_{\bar{l}}^{(g)},\widetilde{\bm{\phi}}_{\bar{l}}^{(g)},\bm{\phi}_{\bar{l} + 1}^{(g)},\widetilde{\bm{\phi}}_{\bar{l} + 1}^{(g)}\mid \mathbf{y}, \mathbf{Z}, \bm{\delta},\beta_{\bar{l}}^{(g-1)},\beta_{\bar{l} + 1}^{(g-1)}\right) = \min\left\{1,\left(\frac{p\left(\bm{\phi}_{\bar{l}+1}^{(g)}\mid \mathbf{y}, \mathbf{Z}, \bm{\delta}\right) \lvert J\left(\widetilde{\bm{\phi}}_{\bar{l}+1}^{(g)}\right) \rvert}{p\left(\bm{\phi}_{\bar{l}}^{(g)}\mid \mathbf{y}, \mathbf{Z}, \bm{\delta}\right) \lvert J\left(\widetilde{\bm{\phi}}_{\bar{l}}^{(g)}\right) \rvert}\right)^{\beta_{\bar{l}}^{(g-1)}-\beta_{\bar{l} + 1}^{(g-1)}}\right\}
                \end{equation*}
                \State Compute 
                \begin{equation*}
                    \rho_{\bar{l}}^{(g)} =  \rho_{\bar{l}}^{(g-1)} + (g+1)^{-\nu}\left(\omega_{\bar{l}}^{(g)}\left(\bm{\phi}_{\bar{l}}^{(g)},\widetilde{\bm{\phi}}_{\bar{l}}^{(g)},\bm{\phi}_{\bar{l} + 1}^{(g)},\widetilde{\bm{\phi}}_{\bar{l} + 1}^{(g)}\mid \mathbf{y}, \mathbf{Z}, \bm{\delta},\beta_{\bar{l}}^{(g-1)},\beta_{\bar{l} + 1}^{(g-1)} \right) - \bar{\omega}\right)
                \end{equation*}
                \State Set $\beta_{0}^{(g)} = 1$ and
                \For{$l=1,\dots,L-1$}
                \begin{equation*}
                    \frac{1}{\beta_{l+1}^{(g)}} = \frac{1}{\beta_{l}^{(g)}} + \exp{\left( \rho_{l}^{(g)} \right)}
                \end{equation*}
                \EndFor
            \EndFor
            \State Return for $l = 1,\dots,L$ draws $\bm{\phi}_{l}^{(g_{0}+1)},\dots,\bm{\phi}_{l}^{(G)}$
        \end{algorithmic}
    \end{algorithm}
    
\section{MCMC results}\label{sec:app_images}
    
    \begin{figure}[htbp!]
        \centering
        \begin{tabular}{l}
            (a) \\
            \includegraphics[width = 0.63\textwidth]{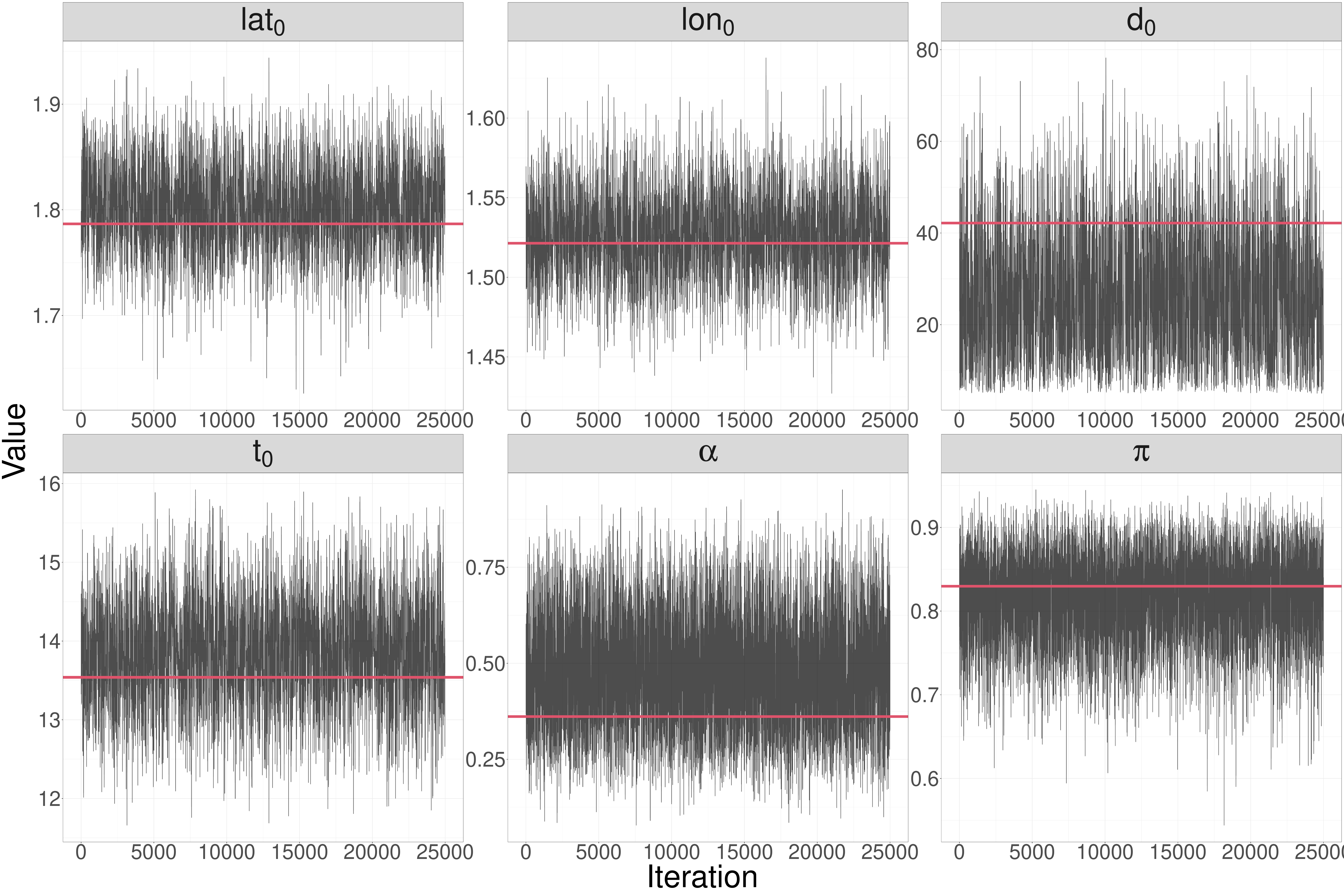} \\
            (b) \\
            \includegraphics[width = 0.63\textwidth]{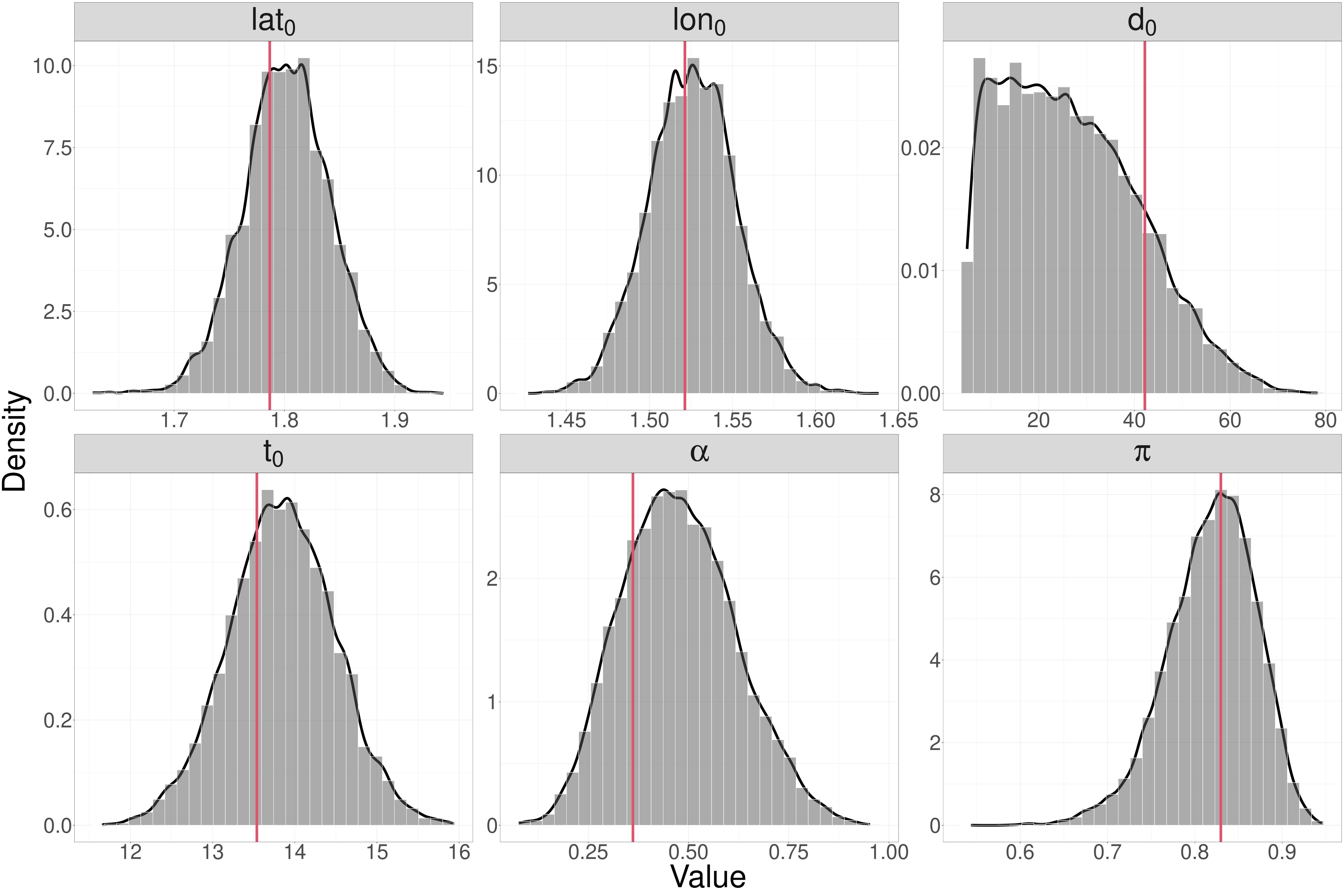} \\
            (c) \\
            \includegraphics[width = 0.63\textwidth]{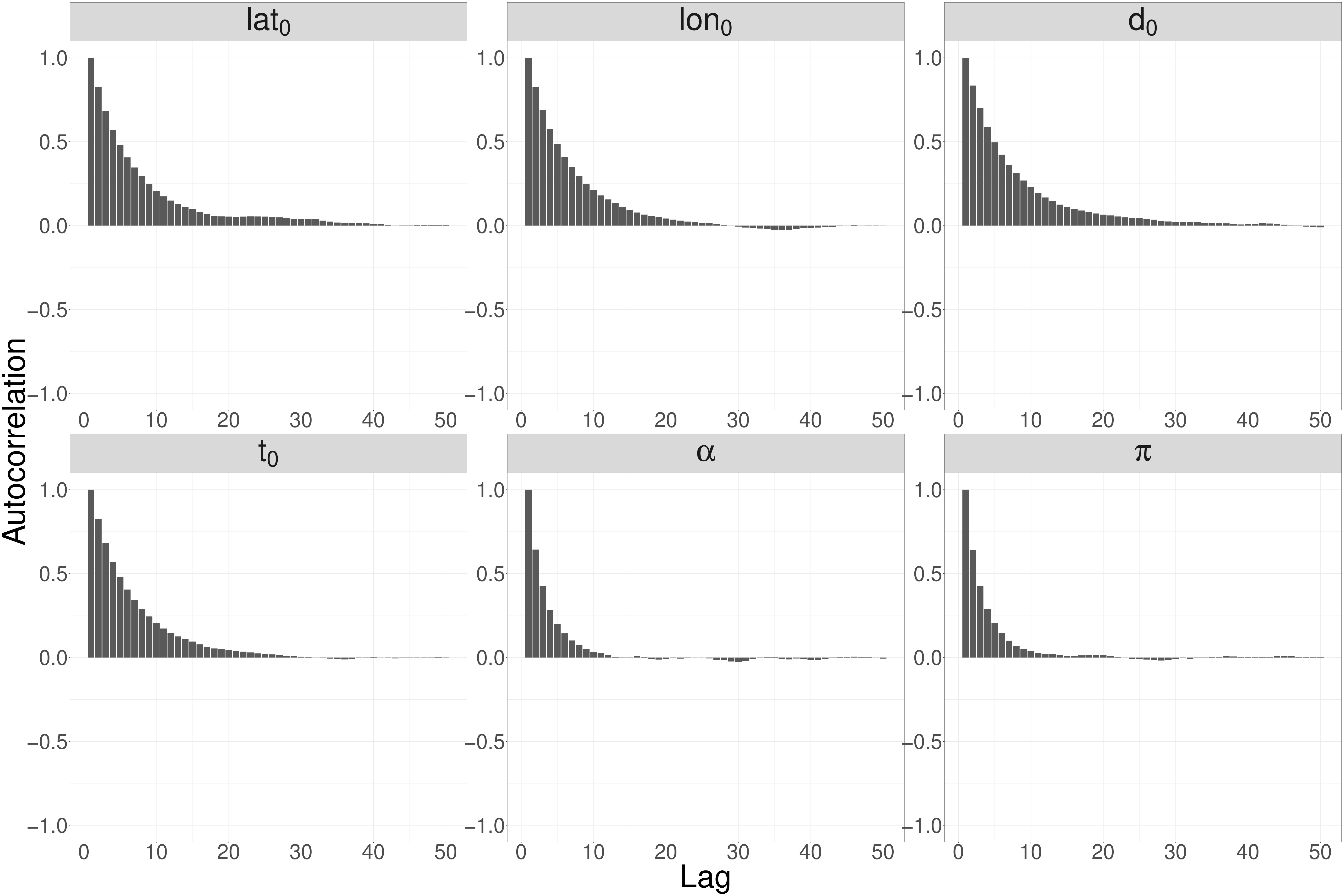}
        \end{tabular}
        \caption{Diagnostics plot for the parameters of a simulation with 100 observations and network N-S and E-W spatial variances $(0.25,1)$. Traceplots (a) densities (b) and autocorrelations (c).}
        \label{fig:sim_diagn}
    \end{figure}

    \begin{figure}[htbp!]
        \centering
        \begin{tabular}{l}
            (a) \\
            \includegraphics[width = 0.66\textwidth]{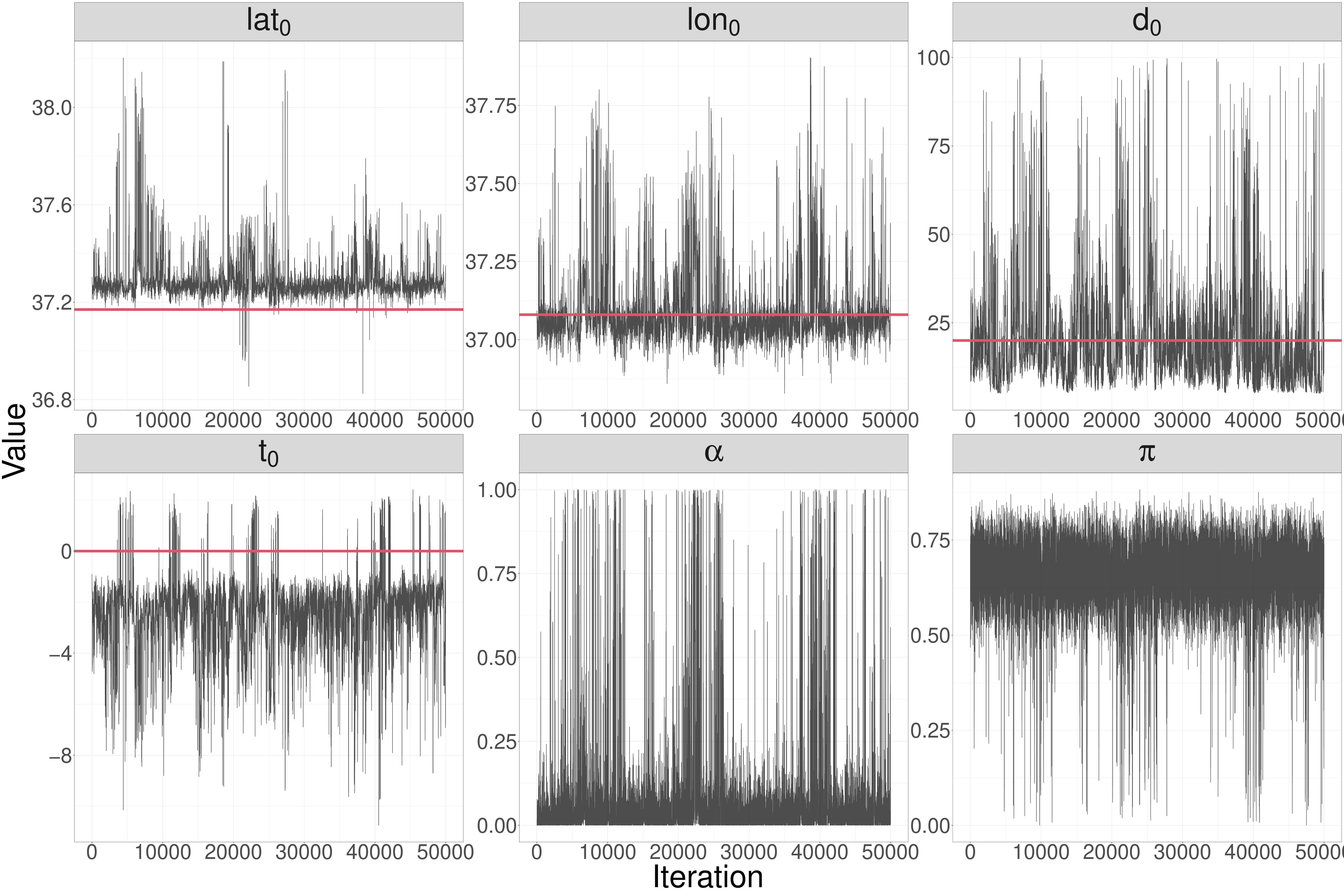} \\
            (b) \\
            \includegraphics[width = 0.66\textwidth]{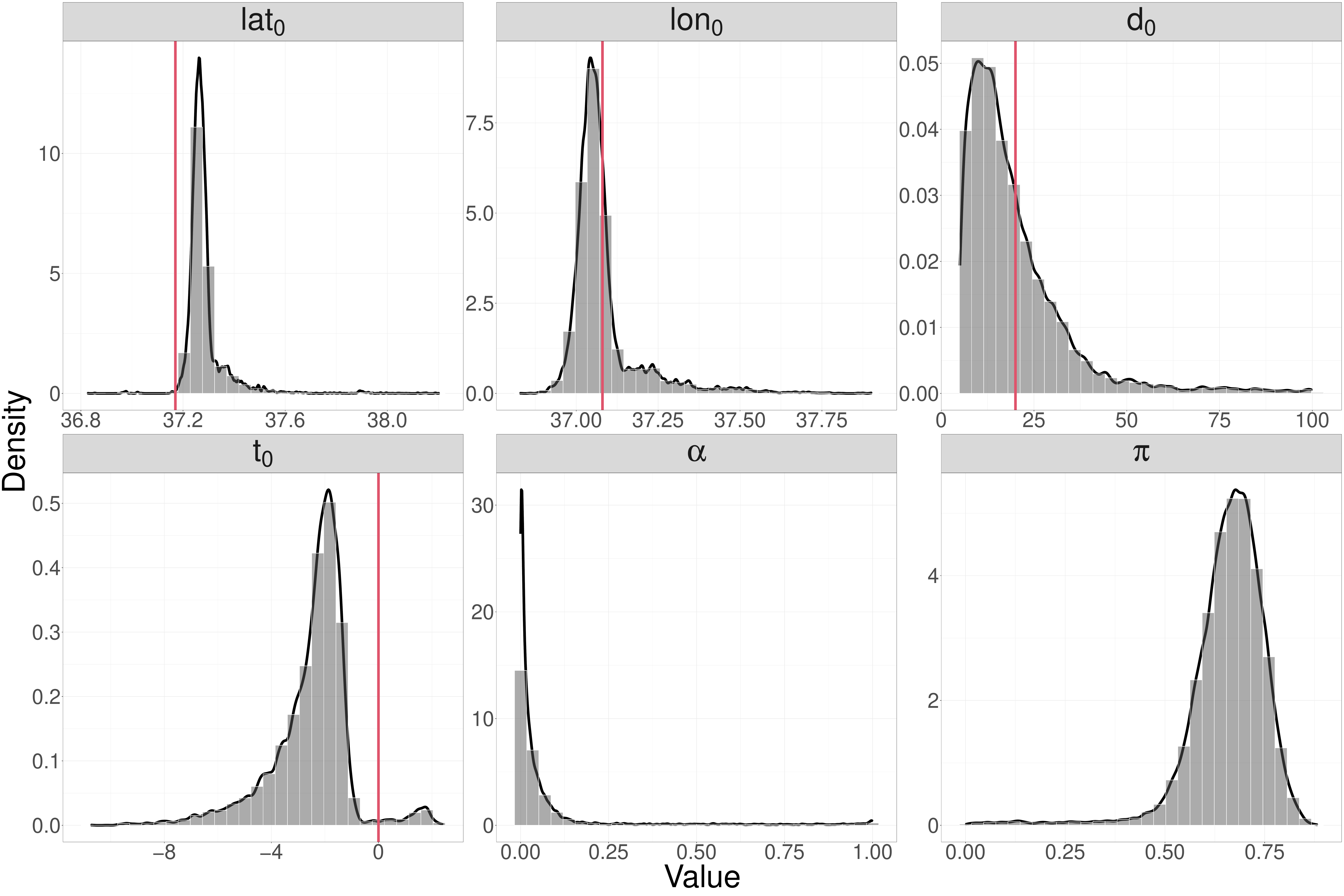} \\
            (c) \\
            \includegraphics[width = 0.66\textwidth]{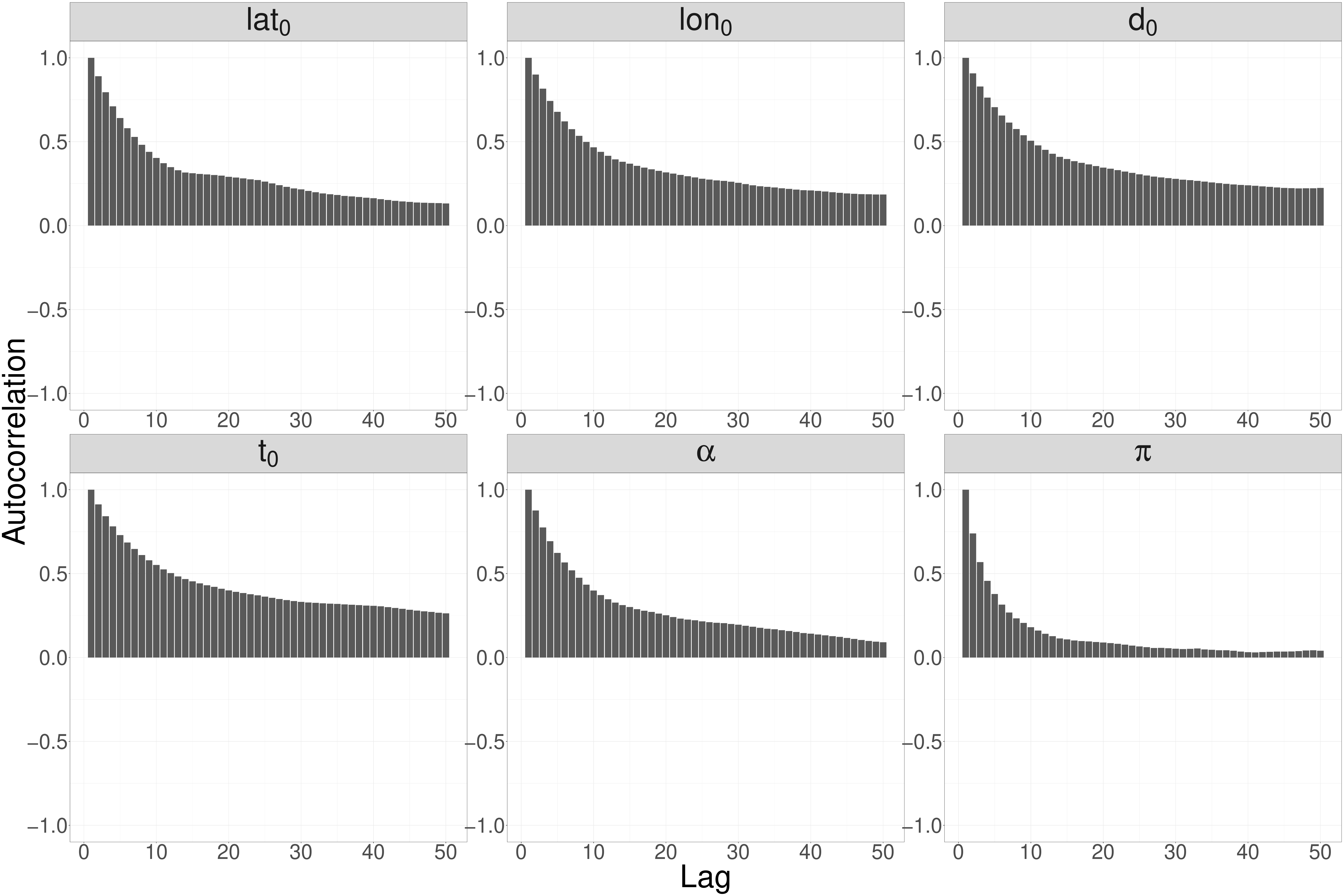}
        \end{tabular}
        \caption{Diagnostics plot for the Turkish-Syrian case. Notice that the reference for $t_0$ is the true earthquake origin time. Traceplots (a), densities (b) and autocorrelations (c).}
        \label{fig:turkey_diagn}
    \end{figure}

    \begin{figure}[htbp!]
        \centering
        \begin{tabular}{l}
            (a) \\
            \includegraphics[width = 0.66\textwidth]{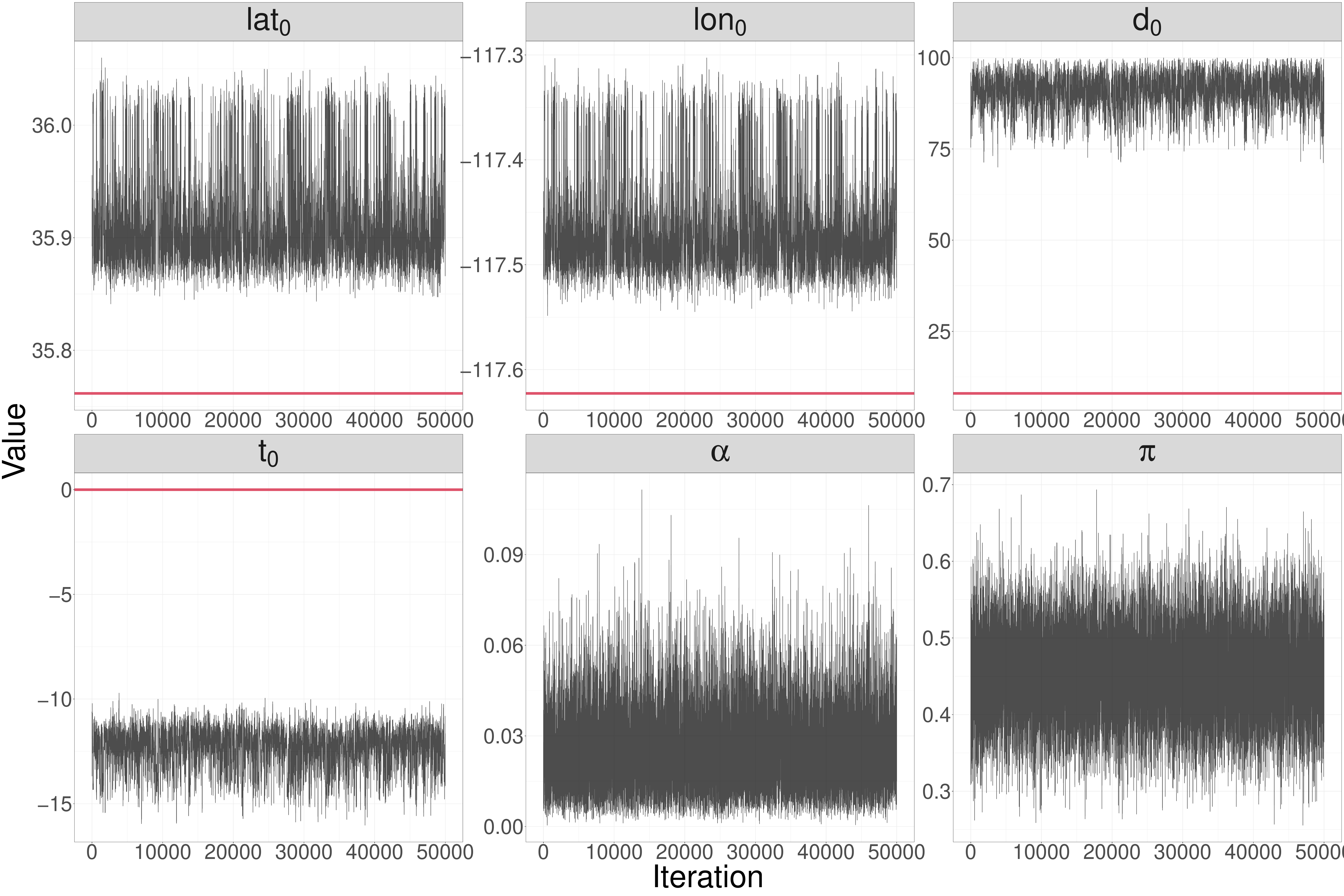} \\
            (b) \\
            \includegraphics[width = 0.66\textwidth]{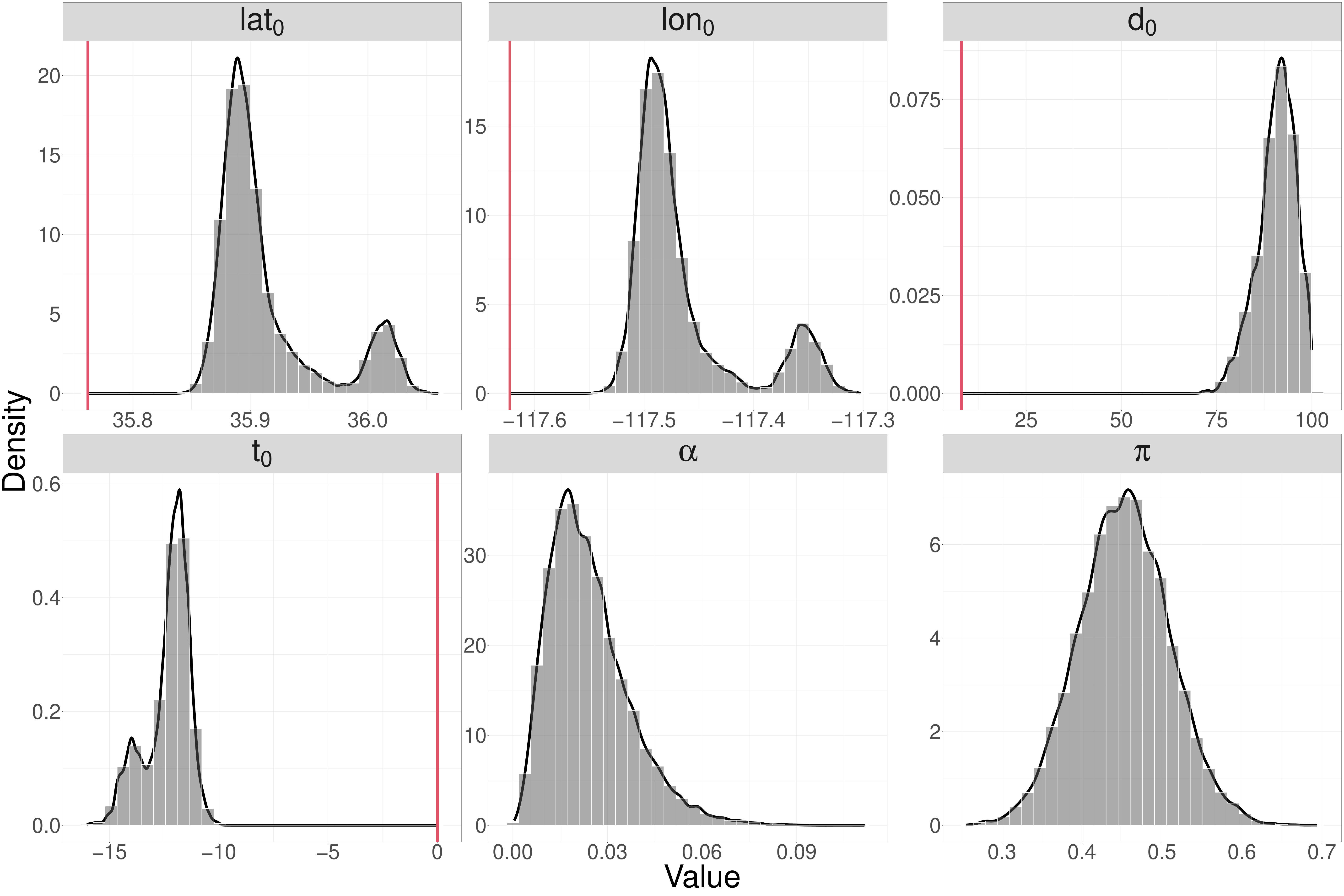} \\
            (c) \\
            \includegraphics[width = 0.66\textwidth]{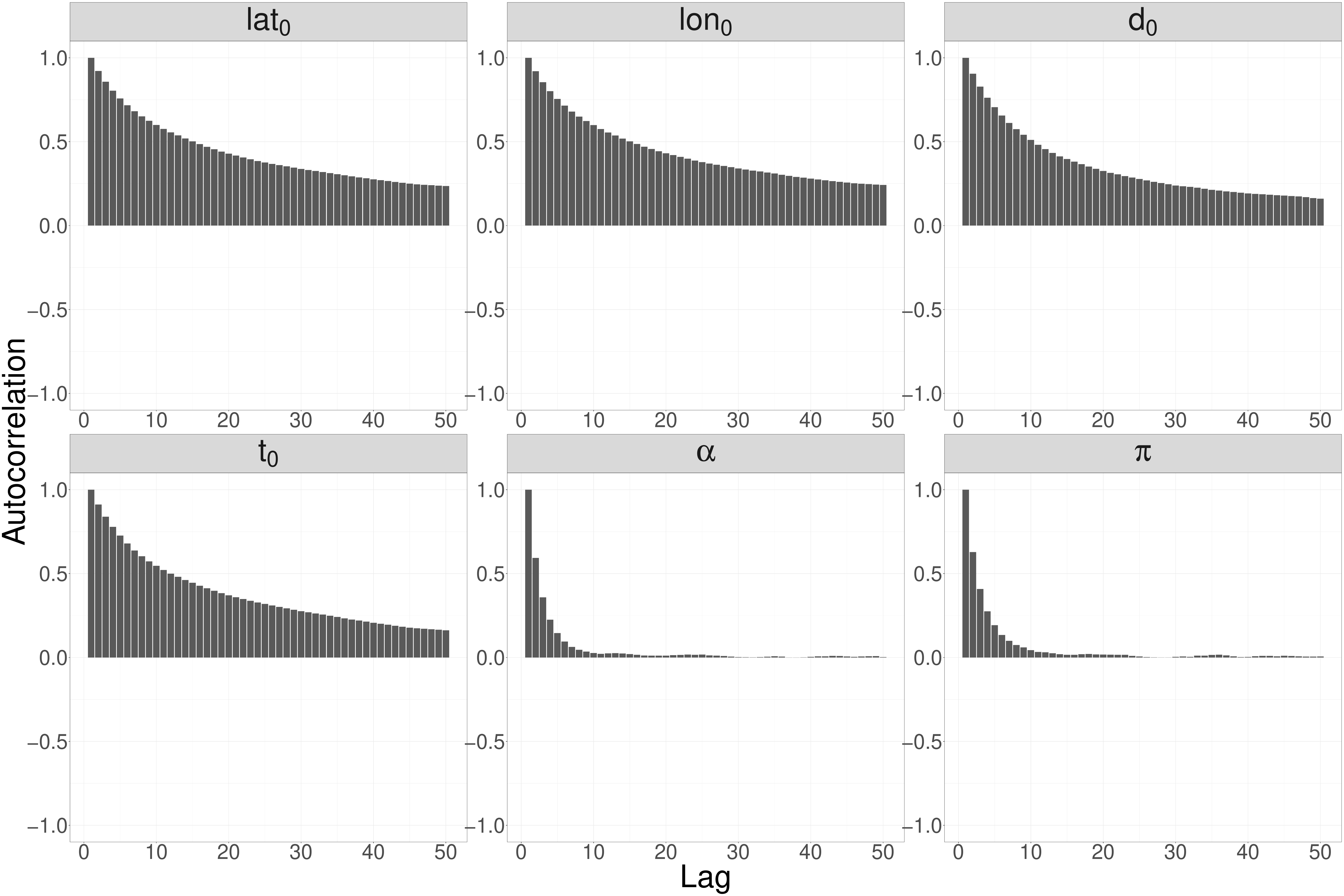}
        \end{tabular}
        \caption{Diagnostics plot for the Californian case. Notice that the reference for $t_0$ is the true earthquake origin time. Traceplots (a), densities (b) and autocorrelations (c).}
        \label{fig:california_diagn}
    \end{figure}

    \begin{figure}[htbp!]
        \centering
        \begin{tabular}{l}
            (a) \\
            \includegraphics[width = 0.66\textwidth]{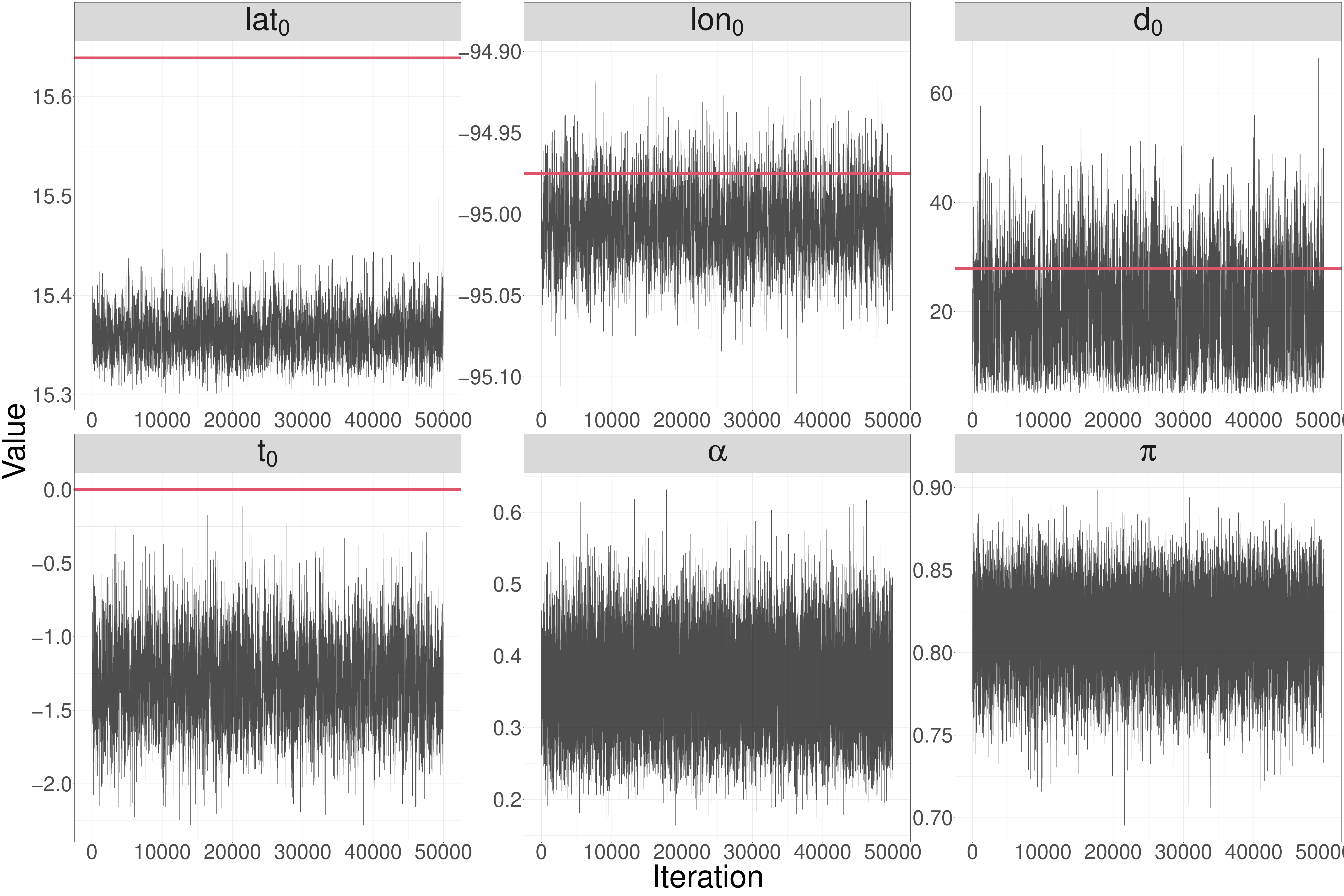} \\
            (b) \\
            \includegraphics[width = 0.66\textwidth]{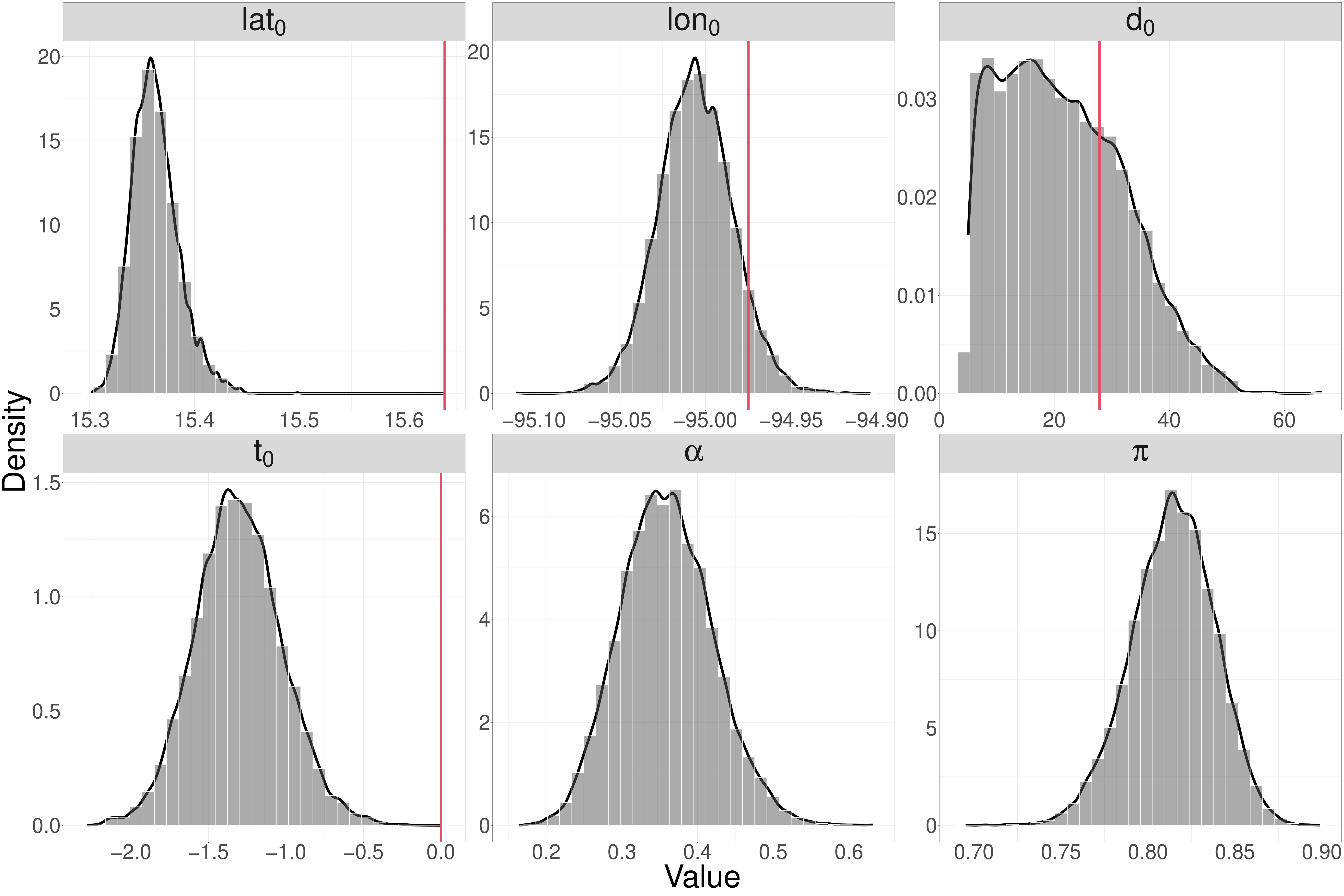} \\
            (c) \\
            \includegraphics[width = 0.66\textwidth]{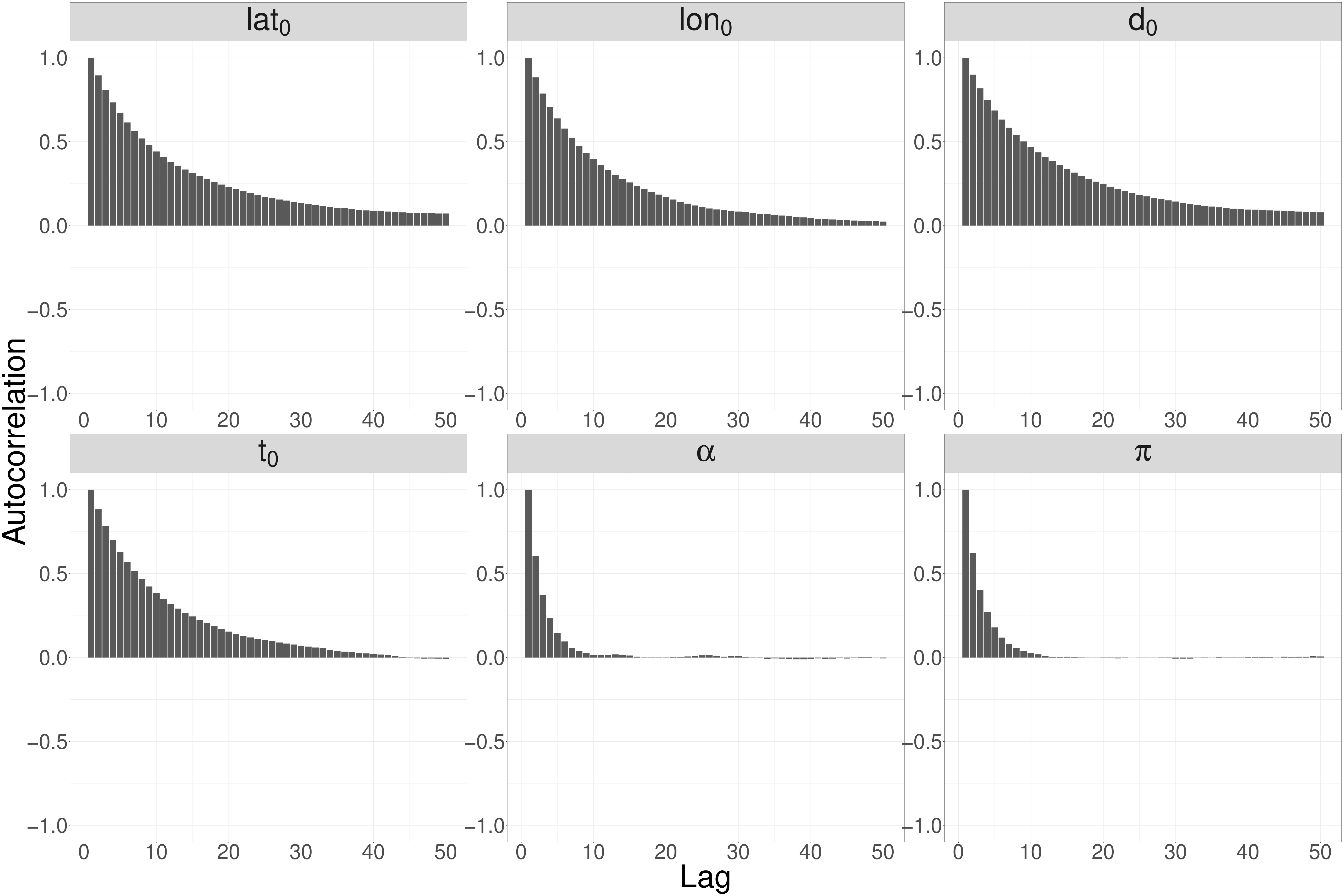}
        \end{tabular}
        \caption{Diagnostics plot for the Mexican case. Notice that the reference for $t_0$ is the true earthquake origin time. Traceplots (a), densities (b) and autocorrelations (c).}
        \label{fig:mexico_diagn}
    \end{figure}
    
\end{appendix}

\newpage

\bibliographystyle{rss}
\bibliography{biblio}

\end{document}